\documentclass{article}
              
\usepackage[a4paper]{geometry}
\usepackage{amsmath,amsfonts,bm,amssymb}
\usepackage{amsthm}
\usepackage{float}
\usepackage{latexsym}
\usepackage{graphicx}
\usepackage[authoryear]{natbib}
\usepackage{rotating}
\usepackage{caption}
\usepackage{epsfig}
\usepackage{color}

\usepackage{pdflscape}

\usepackage[normalem]{ulem}

\newcommand{\luca}[1]{{#1}}

\newcommand{\co}[1]{{#1}}
\newcommand{\red}[1]{{#1}}

\newcommand{\be}{\begin{equation}}
\newcommand{\ee}{\end{equation}}
\newcommand{\ba}{\begin{eqnarray}}
\newcommand{\ea}{\end{eqnarray}}

\theoremstyle{definition}

\theoremstyle{remark}

\usepackage{verbatim}
\usepackage{diagrams}

\newtheoremstyle{commenta}
  {6pt}
  {6pt}
  {\sffamily}
  {}
  {\sffamily \bfseries}
  {:}
  {.5em}
  {}

\theoremstyle{commenta}

\bibpunct{(}{)}{;}{}{,}{,}

\newcommand{\ev}{\mathrm{E}}
\newcommand{\var}{\mathrm{Var}}
\newcommand{\cov}{\mathrm{Cov}}
\newcommand{\cor}{\mathrm{Cor}}
\newcommand{\beq}{\begin{equation}}     
\newcommand{\eeq}{\end{equation}}

\begin{document}

\title{Epistasis and constraints in fitness landscapes}

\author{Luca Ferretti$^{1,2}$\footnote{Email: luca.ferretti@gmail.com}, Daniel Weinreich$^{3}$, Benjamin Schmiegelt$^{4}$, Atsushi Yamauchi$^{5}$, \\ Yutaka Kobayashi$^{6}$, Fumio Tajima$^{7}$ and Guillaume Achaz$^{1}$\footnote{Email: guillaume.achaz@upmc.fr}}
\date{}

\maketitle

(1) Evolution Paris-Seine (UMR 7138) and Atelier de Bio-Informatique, UPMC, Paris; SMILE, CIRB (UMR 7241); Coll\`ege de France, Paris, France.  (2) The Pirbright Institute, Woking, United Kingdom. (3) Ecology and Evolutionary Biology, Brown University, Providence, USA. (4) Institute for Theoretical Physics, University of Cologne, Germany. (5) Center for Ecological Research, Kyoto University, Japan. (6) Kochi University of Technology, Japan. (7) Department of Biological Sciences, University of Tokyo, Japan.

\begin{abstract}
Genotypic fitness landscapes are constructed by assessing the fitness of all possible combinations of a given number of mutations.  In the last years, several experimental fitness landscapes have been completely resolved. \co{As fitness landscapes are high-dimensional, their characterization relies on simple \red{measures of their structure, which can be used as statistics in empirical applications}. Here we propose two new sets of  \co{measures} \co{that explicitly capture} two relevant features of fitness landscapes: epistasis and constraints.} The first set contains new \red{measures } for epistasis based on the correlation of fitness effects of mutations. They have a natural interpretation, capture well the interaction between mutations, can be obtained analytically for most landscape models and can \co{therefore} be used to discriminate between different models. The second set contains measures of evolutionary constraints based on ``chains'' of forced mutations along fitness-increasing paths. Some of these measures are non-monotonic in the amount of epistatic interactions, but have instead a maximum for intermediate values. \co{We further characterize the relationships of these measures to the ones that were previous proposed (e.g. number of peaks, roughness/slope, fraction of non-additive components, etc).} Finally, we show how these  \co{measures can} help uncovering the amount and the nature of epistatic interactions in two experimental landscapes.
\end{abstract}

\section{Introduction}

Fitness landscapes have been a very successful metaphor to study evolution. \red{Most simply}, the idea of Sewall \citet{Wright1932} to view evolution as a hill-climbing process  proved to be appealing and inspired a vast amount of theoretical work in phenotypic and molecular evolution \citep{Orr2005,Visser2014}. Furthermore, this metaphor contributed to the scientific exchange with other fields, especially with computer science \citep{Richter2014} and physics \co{\citep{stein1992spin}}. 

In evolutionary biology, fitness landscapes have been used to study adaptation. In the classical \red{metaphor, an evolving population is abstracted into a particle that navigates }in the landscape \citep{Orr2005}. \red{In a strong selection weak mutation regime \citep{Gillespie1983}, the evolutionary paths followed by the populations are restricted to paths of increasing fitness, which can sometimes even be completely deterministic}. In this perspective, it has been emphasized that many fundamental features of adaptation depend on whether the landscape is smooth or rugged. Among other things, \red{the ruggedness and the properties of fitness landscapes } have been related to speciation processes \citep{Gavrilets2004,Chevin2014}, to the benefits of sexual reproduction \citep{Kondrashov2001,Visser2009,Otto2009,Watson2011}, and more generally, to the repeatability of the adaptation process (e.g. \citet{Kauffman1993,Colegrave2005,Chevin2010,Salverda2011}).

\co{Consequently, it is now clear that several aspects of evolutionary processes directly depend on the structure of the fitness landscapes in which the organisms are evolving. Furthermore, Wright's idea of genotype-fitness landscapes moved from a metaphor to an object of experimental studies, as several fitness landscapes were experimentally resolved \citep{Visser2014}.  In that regard, characterizing the structure of experimental and model fitness landscapes is a key step in our ability to decipher \red{evolution at the finest scale}. However, because fitness landscapes are objects of (very) high dimensionality, their characterization relies on simple \red{scalar }  measures (\textit{i.e.} statistics) that are able to capture the important features of the landscapes. In this study, we propose two new sets of measures for fitness landscapes that have an immediate interpretation in biology in terms of epistasis and evolutionary constraints. }

One of the most basic ingredients \co{that characterize the structure} of fitness landscapes is epistasis. Epistasis is the interaction between the effects of mutations at different loci. It is usually defined as the non-multiplicative part of the fitness effects of combined mutations,  that is the non-additive part, in log-scale.  In \red{the}  presence of epistasis, the fitness effect of a mutation at a given locus depends on the genetic background and \red{consequenty}, a mutation at a given locus changes the distribution of fitness effects of other mutations at other loci. \co{For the 2-loci 2-alleles case, assuming that the genotype with the smallest fitness is labeled $00$, epistasis can be expressed (in logscale) as the departure from additivity : $e = f({11})   + f({00}) - f({10})  - f({01})$, where $f({ij})$ is the malthusian fitness of the genotype $ij$ (e.g. \citet{Phillips2008}).}

Assuming random \red{fitness values } \co{(\textit{i.e.} NK landscapes)}, \citet{Kauffman1993} showed a positive correlation between the amount of epistatic interactions and the ruggedness of a landscape, defined as the density of peaks (genotypes with no fitter neighbors). As the \red{ number of loci that interact together } grows, the landscape is more rugged, and more local peaks end evolutionary paths. \red{At the maximum number of epistatic interactions, } \co{the fitnesses of each genotype are completely uncorrelated, resulting in a so-called House-of-Cards model \citep{Kingman1978,Kauffman1987}, for which several measures on paths of increasing fitness were derived  \citep{Kauffman1987,Franke2011,Hegarty2014,Berestycki2013number}}

Experimental fitness landscapes make possible to test predicted  properties of evolutionary trajectories through evolution experiments, together with sequencing \citep{Achaz2014}. Currently available fitness landscapes are based on mutations in a few loci, typically 4-10 \citep{Szendro2013,Weinreich2013}. Since the number of genotypes scales as the product of the number of alleles at all loci, testing all the combinations of mutations in an organism (or even in a protein) is beyond the reach of any reasonable future experiment. However, small landscapes have been resolved and analyzed. \red{An exciting opportunity raised by } \co{the recent release of these experimental landscapes is to find the adequate model(s) that is (are) able to generate landscapes that share similar features with the observed ones. In that regard}, \co{characterizing the structure} of small fitness landscapes is today a key step for our understanding of evolution in \red{the presence of realistic interactions among mutations}. 

The high dimensionality of fitness landscapes make them almost impossible to visualize (although some attempt were proposed, see \citet{McCandlish2011} \red{or \citet{Magellan}}). As a consequence, the analyses of fitness landscapes will mostly rely on  \co{measures} that capture important features of their structures. Several  \co{measures} were proposed previously and used to analyze experimental fitness landscapes (reviewed in \citet{Szendro2013}). The most natural one that was historically used to appreciate the ruggedness of a landscape is its number of peaks \citep{Weinberger1991}.  Intriguingly, although ruggedness is more adequately represented by both types of extrema, only little attention has been payed to the number of sinks (genotypes with only fitter neighbors). We therefore suggest that peaks and sinks are both adequate measures of the landscape ruggedness. \luca{Most models generate landscapes with the same mean number of peaks and sinks; however, small landscapes can have a different number of peaks and sinks due to random sampling. Furthermore, there is no theoretical reason for a symmetry between peaks and sinks in real fitness landscapes. }

\begin{figure}[ht]
\begin{center}
\includegraphics[scale=0.75]{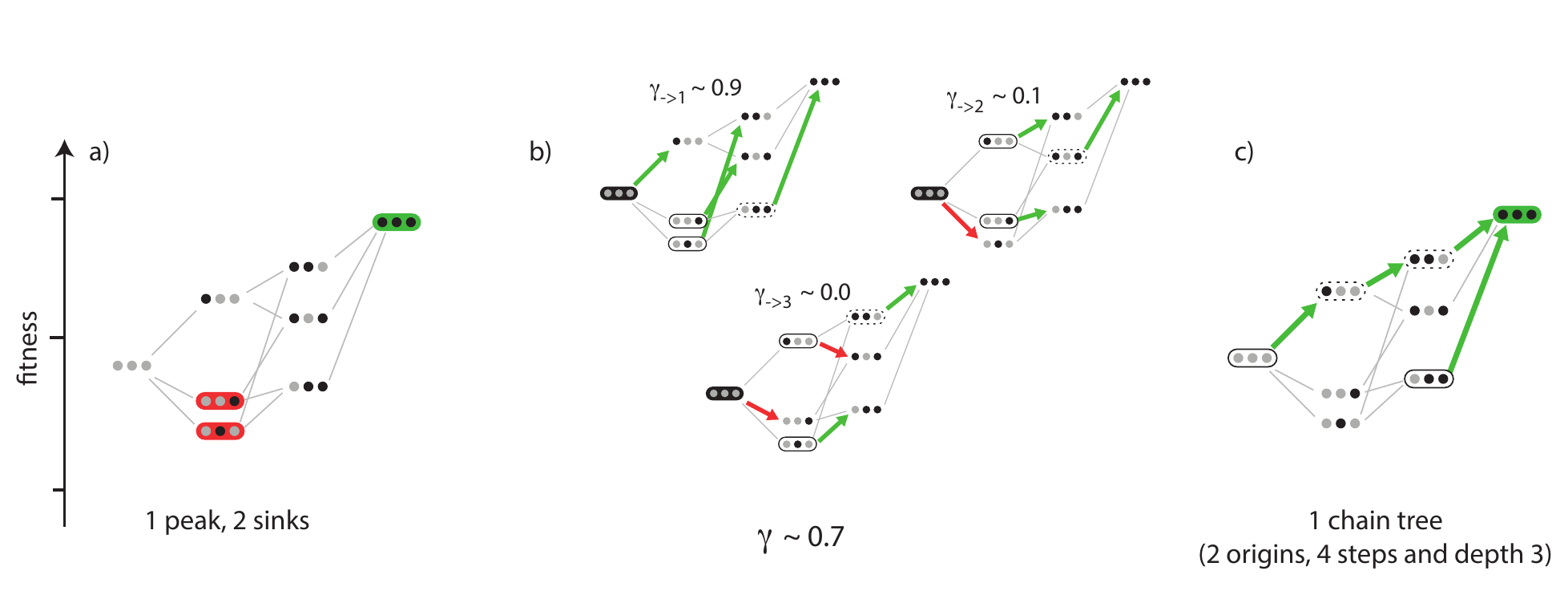}
\caption{ {\bf  \co{Measures} of fitness landscapes}. We depict on the same fitness landscape (of 3 loci with 2 alleles each) three  \co{measures}. \newline
(a) Peaks, here in green, are genotypes with no fitter neighbors whereas sinks are genotypes with only fitter neighbors. \newline
\co{(b) $\gamma$ is the pairwise correlation in fitness effect of mutation between neighbor genotypes. It measures how much another mutation in a genotype affects the focal mutation, averaged across all mutations and the whole landscape. Here the average correlation is good ($\gamma \approx 0.7$). $\gamma_{\rightarrow i }$ is the correlation in fitness effect of mutation $i$ between neighboring genotypes. In the example, mutations at locus 1 are almost independent of the genotypes ($\gamma_{\rightarrow 1} \approx 0.9$), whereas the effects of the mutations at locus 3 show almost no correlationa across genotypes ($\gamma_{\rightarrow 3} \approx 0$)}. \newline
(c) In this landscape, there is a single chain tree that is composed of 4 steps (genotypes with a single fitter neighbor), 2 origins and a depth (the largest number of steps chained together) of 3. }
\label{fig1} 
\end{center}
\end{figure}

\co{Other measures such as $r/s$ ratio \red{(ratio of the roughness over additive fitness)}, fraction of sign epistasis, etc. (see detailed description in the Appendix) were also proposed to characterize the structure of fitness landscapes. As they all represent direct or indirect measures of epistasis, all these measures were shown to be pairwise correlated in experimental fitness landscapes \citep{Szendro2013}. In that regard, other \co{measures} related to evolution but somehow \red{uncorrelated  to } the amount of epistasis are also needed to investigate the \co{nature} of the interactions in the landscapes.

Here, we describe two new \co{measures} that can be used to characterize the structure of fitness landscapes (Figure \ref{fig1}).} The first one, $\gamma$, is the single-step correlation of fitness effects for mutations between neighbor genotypes (Figure \ref{fig1}b). It is a direct measure of epistasis, \textit{i.e.} it measures how much the fitness effect of a mutation is affected when a genotype experiences another mutation. As all correlation measures, $\gamma$ ranges from $-1$ to $+1$ and is a very natural quantity to describe the amount of epistasis (ruggedness) in the landscape. The second one, chains, aims at quantifying the amount of constrained evolution in the landscapes. \red{In the limit of strong selection and low mutation rates, } when a hill climbing evolutionary path reaches a genotype that has a single fitter neighbor, the next step of the path is essentially deterministic. In this case, only one of the mutations available in the landscape can improve the fitness. If several genotypes with a single way out are connected together they form a chain of ``obligatory" mutational events that often ends on a peak (Figure \ref{fig1}c), but could also end on an intermediate genotype.

The two  \co{measures} target different features of the landscapes: $\gamma$ aims at quantifying the amount of epistasis, independently of its nature, whereas $chains$ discriminate between different types of epistasis. \co{In the following, we will present both  \co{measures} in detail, then discuss their relations with the existing \co{measures} and finally quantify them on two experimental landscapes. \red{We also present the mean value of these measures for several landscape models, like the House of Cards (HoC), the Rough Mount Fuji (RMF) and NK landscapes as well as Ising and Eggbox models.} Details on previous \co{measures} and model landscapes used here are given in the Appendix.}

\section{Epistasis as correlation of fitness effects: $\gamma$}

\subsection{Definition}

\begin{figure}[ht]
\begin{center}
\includegraphics[scale=0.75]{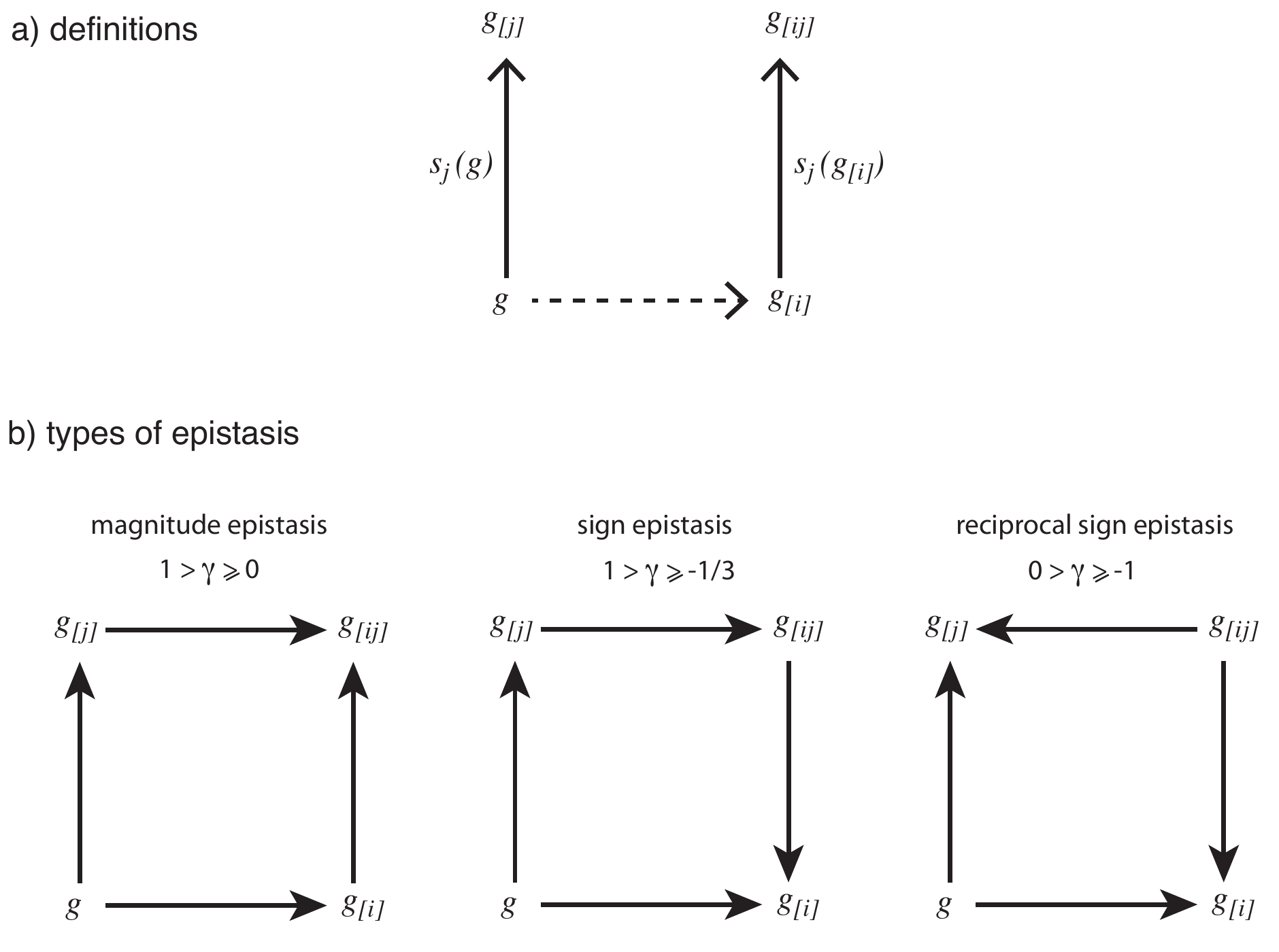}
\caption{  (a) Notation: $\gamma$ is the correlation between the fitness effects $s_j(g)$ and $s_j(g_{[i]})$ over all genotypes $g$ and mutations $i,j$ in the landscape.  (b) Types of epistasis, possible values of $\gamma$ and examples of the corresponding fitness graphs.}
\label{epistatictypes} 
\end{center}
\end{figure}

In this section, we will derive and discuss a new  \co{measure} that is a natural description of the amount of epistasis in fitness landscapes. This new  \co{measure}, denoted by $\gamma$, is simply the correlation of fitness effects of the same mutation in \red{single-mutant } neighbors (see Figure \ref{fig1}b and Figure \ref{epistatictypes}). It measures how the effect of a focal mutation is altered by another mutation at another locus in the background, averaged across the whole landscape.

In the following, we will define it properly in mathematical terms for the bi-allelic case. We denote the (log-scaled) fitness of a genotype $g$ by $f(g)$. \red{We also define }  $g_{[i]}$, the genotype $g$ where the locus $i$ is mutated.  The fitness effect of a mutation at locus $j$, \textit{i.e.} the log-scale selection coefficient of the mutation, is denoted by $s_j(g) = f(g_{[j]})-f(g)$. \luca{The new measure $\gamma$ is then defined as the correlation between two fitness differences $s_j(g)$ and $s_j(g_{[i]})$ measured from genotypes that are one mutation away, as} illustrated by Figure \ref{epistatictypes}a.

\red{Noting that the average of $s_j(g)$ across all genotypes and mutations in the landscape is 0}, we define $\gamma$ as:
\begin{align}
\gamma & =\cor[s (g), s (g_1)]  \label{def_gamma} \\
              & = \frac{ \cov[s(g), s(g_1)]}{\var[s(g)]} \nonumber\\
              &= \frac{\sum_g\sum_i\sum_{j\neq i}s_j(g)\cdot s_j(g_{[i]})}{(L-1)\sum_g\sum_j(s_j(g))^2}\nonumber
\end{align}
where $g_1$ indicates a generic genotype that differs from $g$ by a single mutation. For multiallelic landscapes, the same definition $\gamma=\cor[s(g),s(g_1)]$ can be immediately generalized to any number of alleles.

Even though $\gamma$ is originally defined in term of fitness effects of the mutations, it can be easily recomputed by only using the fitness values themselves. If we denote $\rho_d=\cor[f(g),f(g_d)]$ the correlation between fitness of genotypes apart from distance $d$, it is possible to rewrite $\gamma$ by the simple formula (see proof in the Appendix):
\beq
\gamma = \frac{\rho_1-\rho_2}{1-\rho_1}\label{eqrho}
\eeq
Besides its general interest, this formulation allows \red{us } to measure $\gamma$ in the presence of missing data. Indeed, fitness correlation functions \red{do not need fitness data for all combinations of some set of mutations in order to be estimated.}

\subsection{Interpretation}
To make clear that the above  \co{measure} is a metric of epistasis, we rewrite the above equation as
\begin{align}
\gamma & =1-\frac{\ev[(s(g)-s(g_1))^2]}{2\ev[s^2]} \label{def_gamma2}\\
              & = 1-\frac{\ev[e^2]}{2\ev[s^2]} \nonumber \\
              & =1-\frac{\sum_g\sum_i\sum_{j\neq i}(s_j(g)- s_j(g_{[i]}))^2}{2(L-1)\sum_g\sum_j(s_j(g))^2} \nonumber 
\end{align}

When there is no epistasis, the fitness effects do not depend on the background and $\gamma=1$, \textit{i.e.} perfect correlation between fitness effects. The deviation of $\gamma$ from 1 is proportional to the square of $e_{ij}=f(g_{[ij]})-f(g_{[i]})-f(g_{[j]})+f(g)$, which is a standard measure of the amount of epistatic effect (for 2-alleles 2-loci), normalized by the average squared fitness effect. Thanks to its normalization, this measure of epistasis is not affected by the scale and the absolute level of fitness, but only by relative differences in fitness. \co{Shifting fitnesses by a multiplicative or additive factor does not change this measure.} 

The  \co{measure} $\gamma$ is \co{defined as a correlation}, therefore it is bounded by $-1\leq \gamma\leq 1$, with $\gamma=1$ in the case of no epistasis. The value of $\gamma$ is related to the prevalent type of epistatic interactions (see proof in the Appendix), which are summarized in Figure \ref{epistatictypes}b: 
\begin{itemize}
 \item \red{magnitude epistasis refers to pairwise interactions that do not change the signs of fitness effects. Magnitude epistasis } would still result in a positive correlation between fitness effects, therefore $\gamma$ would still be positive even if smaller than 1: $1> \gamma \geq 0$; 
 \item \red{sign epistasis refers to pairwise interactions where the fitness effects of one mutation change sign after the other mutation. Sign epistasis } would contribute with terms of both signs to the correlation, therefore resulting in values centered around 0: $1> \gamma \geq -1/3$; 
 \item finally, \red{reciprocal sign epistasis refers to pairwise interactions where both fitness effects change sign. Reciprocal sign epistasis } would imply a negative correlation between fitness effects, and therefore a negative value of $\gamma$: $0> \gamma \geq-1$. 
\end{itemize}

The deviation of the mean value of $\gamma$ from 1 for simple landscape models measures epistasis as a function of the parameters of the models (see the Appendix for details on the derivations\red{; the reader unfamiliar with the models will also find a brief presentation there}). For example, in NK landscapes the epistasis grows with \red{the parameter K describing the number of loci involved in each interaction } and in fact we have \red{the approximate equation}
\beq
\ev[\gamma] \simeq 1-\frac{K}{L-1}.\label{eqnk}
\eeq
For the HoC model\red{, i.e. a maximally uncorrelated landscape, } we have $K=L-1$ and therefore
\beq
\ev[ \gamma]\simeq 0,
\eeq
\textit{i.e.} this model shows strong random epistasis.

For RMF models, which are combinations of an additive landscape \red{and a completely uncorrelated one, } the correlation of fitness effects is 
\beq
\ev[\gamma] \simeq 1-\frac{2\sigma_{HoC}^2}{\mu_a^2+\sigma_a^2+2\sigma_{HoC}^2}\label{eqhoc}
\eeq
\red{where $\mu_a$ and $\sigma_a$ are the mean and variance of the additive fitness effects and $\sigma_{HoC}^2$ is the variance of the uncorrelated HoC component. Therefore, in this case, the measure of epistasis is proportional to the variance contribution of the uncorrelated component.}

\subsection{Epistasis for specific mutations}

The correlation of fitness effects is also a useful measure of the interaction between specific mutations. Some simple generalizations of the $\gamma$  \co{measure} are:
\begin{itemize}
 \item 
$\gamma_{i\to}$, which describes the epistatic effect of a mutation in \co{locus} $i$ on other \co{loci}: 
\beq
\gamma_{i\to}=\cor[ s(g),s(g_{[i]}) ]=\frac{\sum_g\sum_{j\neq i}s_j(g)\cdot s_j(g_{[i]})}{\sum_g\sum_{j\neq i}(s_j(g))^2} ,
\eeq
\item $\gamma_{\to j}$, which describes the epistatic effects of other mutations on \co{locus} $j$: 
\beq
\gamma_{\to j}=\cor[ s_j(g),s_j(g_1) ]=\frac{\sum_g\sum_{i\neq j}s_j(g)\cdot s_j(g_{[i]})}{(L-1)\sum_g(s_j(g))^2}
\eeq
\item $\gamma_{i \to j}$, which is a matrix that describes the epistatic effect of \co{locus} $i$ on \co{locus} $j$: 
\beq
\gamma_{i \to j}=\cor[s_j(g),s_j(g_{[i]})]=\frac{\sum_gs_j(g)\cdot s_j(g_{[i]})}{\sum_g(s_j(g))^2}
\eeq
\end{itemize}
These measures can \red{also be } generalized easily to multiallelic landscapes by considering pairs of mutations at different \co{loci}.

The  \co{measure} $\gamma_{i\rightarrow}$, $\gamma_{\rightarrow j}$ and especially $\gamma_{i\rightarrow j}$ are useful for exploratory and illustrative purposes, since they summarize the interactions between mutations in a clear and compact way, as it can be seen in Figure \ref{fig_gammaij}.

It is also possible to use the more direct measure $\ev[e_{ij}^2]$ \red{as an }  alternative to $\gamma_{i \to j}$. The difference lies in the normalization: $\gamma_{i \to j}=1-\ev[e_{ij}^2]/2\ev[s_j^2]$, therefore $\gamma_{i \to j}$ treats both large and small mutations in the same way while $\ev[e_{ij}^2]$ is larger for large mutations. The choice of the most appropriate measure depends on the question\red{, i.e. if the focus is on the interactions across all mutations, or only the largest ones}. 

\subsection{Decay of the correlation with distance}

The $\gamma$  \co{measure} \red{provides information on the amount of epistasis } but cannot discriminate between different types or models of fitness landscapes, as it occurs for any \red{single } measure of the amount/strength of epistatic interactions. In fact, there are many landscapes with widely different structure but with the same $\gamma$. For example, a HoC model realization would have $\gamma=0$ as \red{would } a landscape composed by an equal mixture of additive and \red{reciprocal sign } epistatic interactions (like in an EMF model). 

However, a natural and interesting extension of this  \co{measure} is given by the full decay of the correlation of fitness effects with distance $d$\red{, which correspond to the cumulative epistatic effect of $d$ mutations } and can be defined as:
\begin{align}
 \gamma_d &=\cor[s(g),s(g_d)]\label{eq_gammad}\\ 
&=\frac{\sum_g\sum_{i_1}\sum_{i_2>i_1}\ldots \sum_{i_d>i_{d-1}}\sum_{j\neq i_1,i_2\ldots i_d}s_j(g)\cdot s_j(g_{[i_1i_2\ldots i_d]})}{{L-1 \choose d} \sum_g\sum_j(s_j(g))^2}\nonumber
\end{align}
where $\gamma_1=\gamma$. \red{As with } $\gamma$, $\gamma_d$ can be expressed in terms of the fitness correlation functions at distance $d$, $\rho_d$:
\beq
\gamma_d=\frac{\rho_d-\rho_{d+1}}{1-\rho_1}\label{eq_rhod}
\eeq

\begin{figure}
\begin{center}
\includegraphics[scale=0.7]{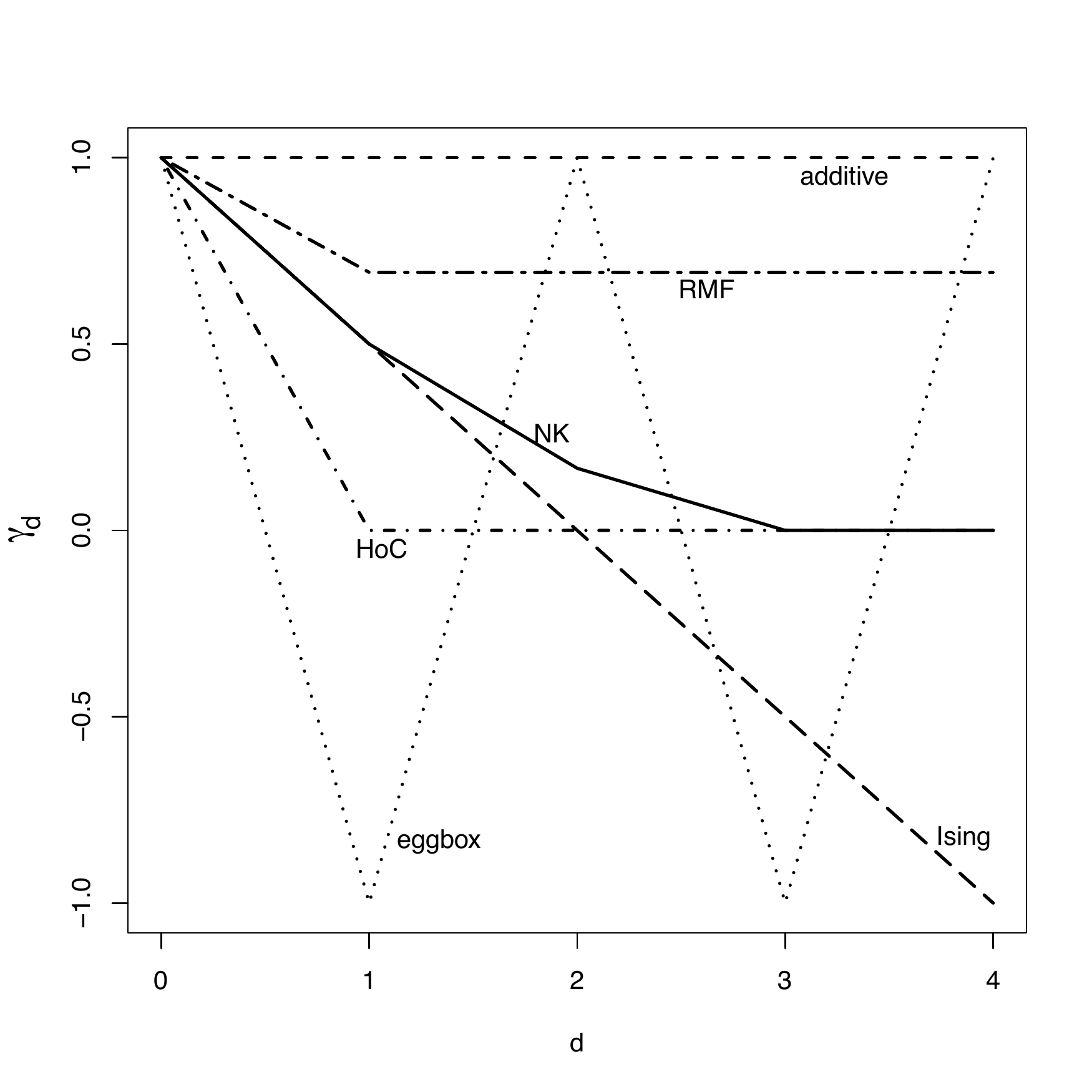}
\caption{Behaviour of the average correlation of fitness effects $\gamma_d$ at different distances in model landscapes with $L=5$. The NK landscape has $K=2$ and the RMF is a mixture of 60\% additive component and 40\% HoC. Analytical formulae are given in the Appendix.}\label{fig_gammad} 
\end{center}
\end{figure}

The decay of $\gamma_d$ with the Hamming distance $d$ is an interesting object of study in itself \red{, since it describes how the epistatic effects of different mutations interact with each other and their cumulative effect. The mean of $\gamma_d$ } can be computed analytically in most models of fitness landscapes - see the first section of the Appendix - and it brings extra information on the structure of the landscape. Different models have a different behaviour (Figure  \ref{fig_gammad}): RMF and HoC models show an abrupt fall already at $d=1$ and then a flat profile, while NK models have a gradual, approximately exponential decay with rate $K/(L-1)$ (Supplementary Figure 1). Models based on The Ising model \red{(based on pairwise reciprocal sign epistasis) } decays linearly until $-1$, while the eggbox \red{(maximally epistatic, anticorrelated) } oscillates between $-1$ and $1$.

\co{Note that the fact that $\gamma^2\leq 1$ \red{together with equation (\ref{eq_rhod}) implies } that there is a general bound on the decay of the fitness correlation functions with distance. In detail, the bound is
\beq
|\rho_{d+1}-\rho_d|\leq 1-\rho_1 
\eeq
\textit{i.e.}, the decay of fitness correlation functions is bounded by the 1-step correlation function.}

\subsection{Correlation in signs  ($\gamma^*$)}

In many experimental situations, fitness is not clearly measurable on an absolute scale, but it is possible to rank the genotypes in order of increasing fitness, or at least to state if a mutation is deleterious or beneficial.

The fitness landscape can be then represented as an acyclic oriented graph, \textit{i.e.} an oriented network where links between genotypes represent single, fitness-increasing mutations. Hereafter, we will refer to this graph as the fitness graph. As an example, the fitness graphs corresponding to different types of epistasis for 2 loci are \co{illustrated in Figure \ref{epistatictypes}b.}

In this context, it is still possible to measure epistasis via the same method by employing a modified  \co{measure} $\gamma^*$ which uses just the sign of the fitness effects, instead of their value. \red{We define $s_j^*(g)$ as the sign of $s_j(g)$. A more robust variant would be}
\beq
s_j^*(g)=\begin{cases}
             +1 \quad \mathrm{for} \quad s_j(g)>\epsilon \\ 
             0 \quad \mathrm{for} \quad -\epsilon\leq s_j(g)\leq \epsilon \\ 
             -1 \quad \mathrm{for} \quad s_j(g) <-\epsilon  
             \end{cases}\eeq 
where $\epsilon$ is a tolerance parameter (possibly depending on the genotype, and larger than the experimental errors). The  \co{measure} $\gamma^*$ is defined as before:
\beq
\gamma^*=\cor[s^*(g),s^*(g_1)]=\frac{\sum_g\sum_i\sum_{j\neq i}s_j^*(g)\cdot s_j^*(g_{[i]})}{(L-1)\sum_g\sum_j(s_j^*(g))^2}\label{def_gammastar}
\eeq

If the landscape has no \red{neutral mutations, we can show that } this  \co{measure} is related to other commonly employed  \co{measures} for fitness graphs. Consider all possible \red{pairwise mutational } motifs in the fitness graph and classify the type of epistasis in each motif as magnitude epistasis, sign epistasis and reciprocal sign epistasis \red{(see Figure \ref{epistatictypes}b)}. We denote the fraction of motifs in each class \co{by $\phi_m$, $\phi_s$ and $\phi_{rs}$ respectively. We have the relation (see proof in the Appendix)
 \beq
\gamma^*=1-\phi_s-2\phi_{rs}\label{gammamotifs}
\eeq
}

What is even more interesting is that both in models and in real landscapes, the results of $\gamma$ and $\gamma^*$ are often \red{numerically } close and highly correlated \red{(see below)}. The only exception is represented by landscapes with weak epistatic interactions dominated by magnitude epistasis, \red{where $\gamma^*=1$. This suggests that $\gamma^*$ could be used in place of $\gamma$ for landscapes where only fitness ranks are known. } These  \co{measures} represent therefore a bridge between fitness graph-based measures and quantitative measures based on absolute fitness.

\section{Constraints in mutation order: {\em chains}}
Many fitness landscape  \co{measures} are correlated with the amount \red{and } strength of epistasis in the landscape. However,  landscapes with similar epistasis could have widely different \red{structure of epistatic interactions}, origin and evolutionary properties. One of these properties \red{is the amount of } constraints on the possible evolutionary paths.

A natural measure of evolutionary constraints is given by the abundance and structure of \red{maximally } constrained paths. Our aim is to characterize these paths in a simple and effective way. For that, we focus on genotypes \red{with only a single beneficial mutation}. All fitness-increasing paths \red{that pass through such genotype } will share this mutation. Some landscape have ``chains'' of consecutive mutations with this properties (see Figure \ref{fig1}c). We \red{now examine } the abundance and size of these chains as measures of evolutionary constraints.

We define a \emph{chain step} as a mutation $g\rightarrow g'$ that is the only possible fitness-increasing mutation from the genotype $g$. \red{Chain steps can occur one after another, forming a linear path of obligatory mutational steps, that we call a $chain$. Several chain steps can lead to the same genotype, but a genotype can have at most one outgoing chain step}. For this reason,  chains can form tree-like structures, that we call a \red{\emph{chain tree}.  A chain tree} is formally the set of all genotypes which are forced to evolve along obligatory paths up to a common final genotype, \textit{i.e.} a maximal groups of connected chain steps.  This definition implicitly assumes a strong selection regime \citep{Gillespie1983}, as there is no fixation for mutations of negative fitness effect. In an additive landscape, there is a single \red{chain tree} containing \red{only those} genotypes one mutation away from the \red{(single)} peak. 
Note that other landscapes can contain more than one \red{chain tree} (or none). We compute also the number of \emph{origins}, that are all genotype that are initial points of a \red{chain} (that obviously excludes intermediate steps).  And finally, we \red{compute also} the maximal \emph{depth} of all chains in the landscape, that is, the maximum number of consecutive steps. In an additive landscape, the depth of the only \red{chain tree} is 1.

\subsection{Out-degree distribution of fitness graphs}

Chains are a natural choice for a measure of evolutionary constraints in the framework of fitness graphs. In particular, \co{chain steps} are strongly related to the out-degree distribution of the fitness graph corresponding to the landscape. 

In fact, there is a one-to-one correspondence between chain steps and genotypes with a single fitness-increasing mutation. The number of fitness-increasing mutations from a given genotype is the out-degree of the genotype in the fitness graph. Therefore, chain steps are simply nodes of out-degree 1.

One of the most well-studied  \co{measures} of fitness landscapes is the number of peaks. Since there are no fitness-increasing mutations from a peak, peaks are simply nodes with out-degree 0.  Therefore, the number of peaks is actually the first bin of the out-degree distribution. Similarly, the number of sinks correspond to the number of genotypes with out-degree $L$ (the last bin of the out-degree distribution). The number of chain steps is the second bin of the out-degree distribution of the fitness graph, and is therefore a natural step further in the characterization of the distribution of out-degree.

For the small empirical landscapes currently available ($L=4-10$), chains contain most of the local information about evolutionary constraints. For larger landscapes, nodes with out-degree 2, 3, or more would also be relevant to assess the amount of constraints. For these landscapes, the full out-degree distribution could be an interesting object of study.

\begin{figure}[ht]
\begin{center}
\includegraphics[width=0.6\textwidth]{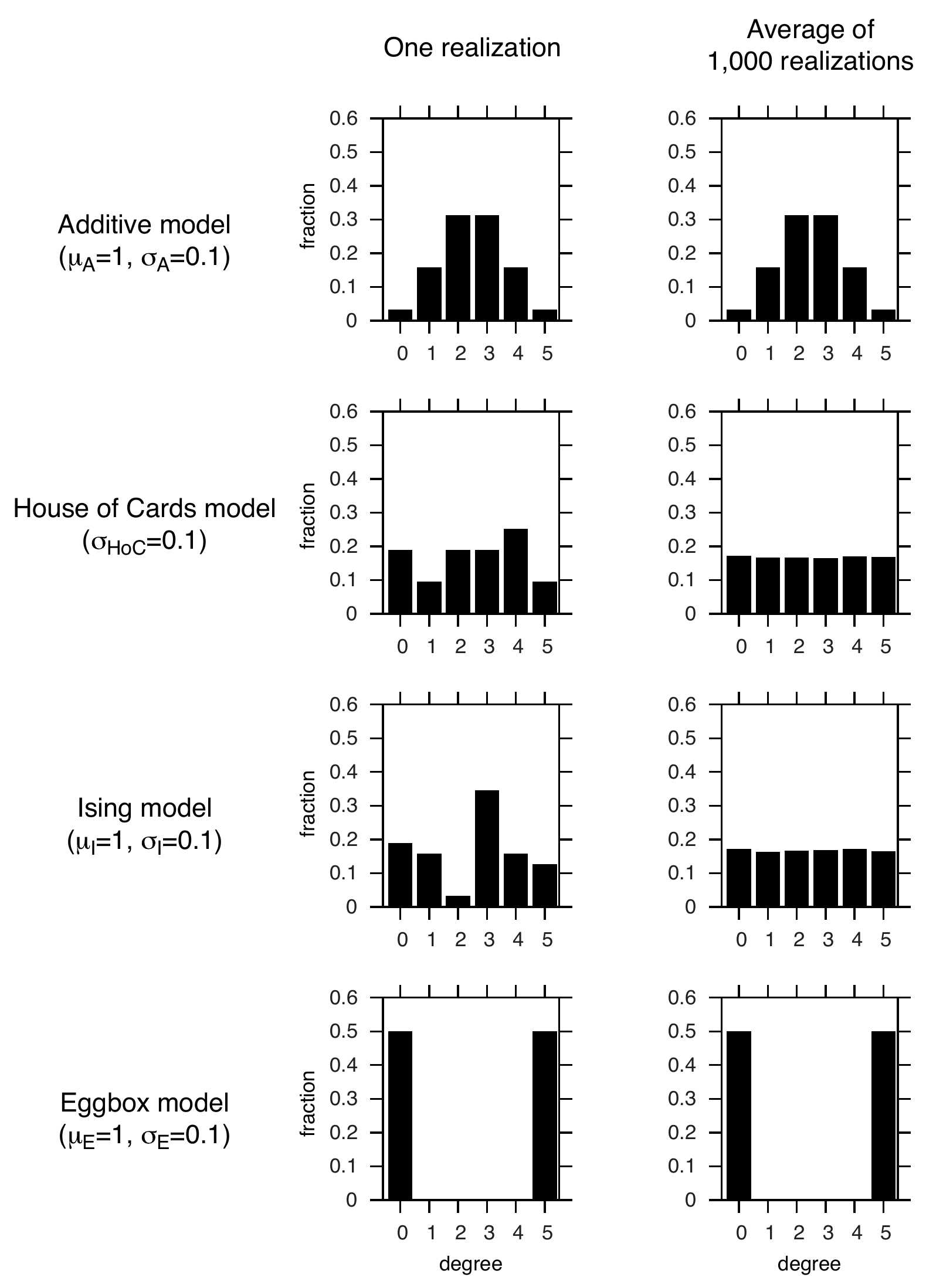}
\caption{ \co{{\bf Out-degree distributions for model landscapes}. The distributions of out-degree (number of fitter neighbors for all loci of the landscape) for four different types of interactions. On the left panel, we report results for a single realization whereas, the right panel reports averages over 1,000 replicates.}}
\label{outdegree} 
\end{center}
\end{figure}

\red{In Figure \ref{outdegree}, are reported the out-degree distributions for four types of epistatic interactions (none, random, pairwise incompatibilities and compensatory at higher order) that correspond to four different theoretical landscape models (Additive, HoC, Ising and Eggbox). Results show that the out-degree distribution usually differs for different types of interactions. However, it is noteworthy to mention that Ising and HoC models show the same average out-degree distributions even though the nature of their epistatic interactions are fundamentally different as well as their overall structure (see Figure \ref{models} and details of the models in the Appendix). }

\co{\red{Interestingly}, only the number of chain steps is contained in the \red{out-degree} distribution. Indeed neither chain depth nor the number of chain origins can be computed from \red{just} the \red{out-degree} distribution. Both values really depend on the structure of the \red{chain tree} itself. A star-like tree, as we expect in additive landscapes, will have several origins and only a depth of 1, whereas a deep chain with a single origin is expected under an IMF model \red{(Ising Mount Fuji is a mixture of additive with Ising pairwise interactions, see details of the models in the Appendix)}. In this last case, the chain corresponds to the subsequent replacements of alleles at neighboring loci starting from one edge up to the other. Chain depth and the number of origins  somehow relate to correlations in the outdegree of nodes in the fitness increasing paths. \red{This stresses the interest of studying both the out-degree distribution and the chain trees in fitness landscapes.}}

\subsection{Connection between chains and the amount of epistasis}

In an additive landscape, all the $L$ genotypes around the peak are origins of a \red{chain tree} that ends at the peak. This starlike \red{chain tree} of depth 1 around the peak is the only \red{chain tree} in these landscapes. In contrast, in a HoC model, there are many small and slightly deeper \red{chain trees}; their number is of the order of $2^L/(L+1)$ (see proof in the Appendix). Therefore, it would be \co{tempting to conclude} that \red{the number of chain trees (as well as other measures of chains)} are correlated to \red{the amount of} epistasis. However, this is not the case.

\red{For example,} in an RMF model with equal additive contribution, it is possible to derive exact theoretical results for the mean of some \red{chain tree}  \co{measures} --most relevant, the number of chain steps. Equations are presented in appendix B. \red{The results are illustrated in Figure \ref{chain_stats}, where the mean of several chains measures are reported as a fraction of additive component in the RMF model. Because the HoC component is fixed (fixed $\sigma_{HoC}$), the higher $\mu_a$, the more additive is the model ($\mu_a$ is the mean additive effect of mutations) ; consequently, when $\mu_a \to \infty$, the model becomes additive whereas it converges to an HoC model when $\mu_a \to 0$.}

\begin{figure}
\includegraphics[width=\textwidth]{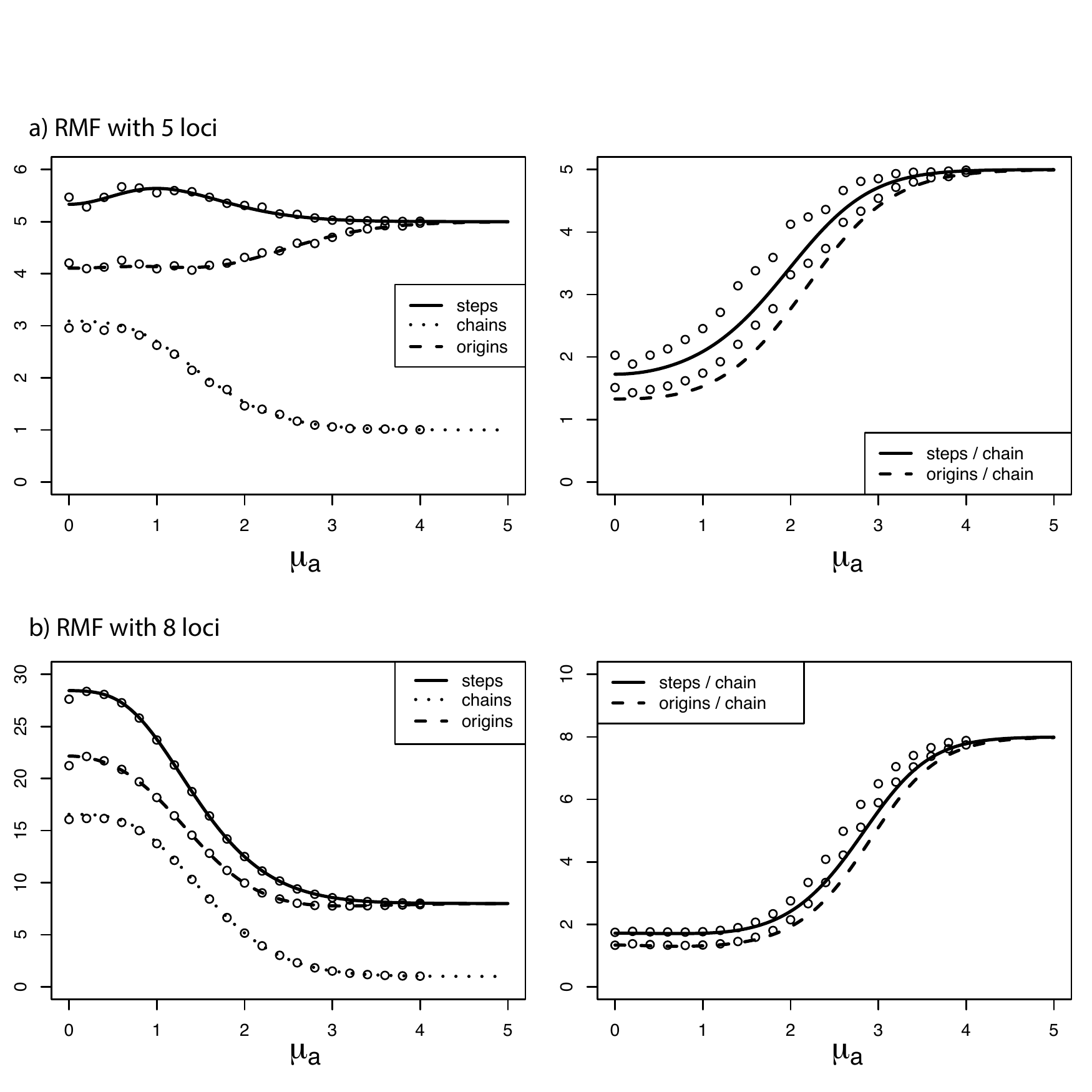}
\caption{Chain  \co{measures} (number of steps, number of chains, number of origins) as a function of the additive fitness effect $\mu_a$ for a RMF landscape with $\sigma_a^2=0$, $\sigma_{HoC}^2=1$. Lines represent the analytical mean values, while dots represent the average over $10^4$ simulations.} \label{chain_stats}
\end{figure}

\begin{figure}[ht]
\begin{center}
\includegraphics[scale=0.5]{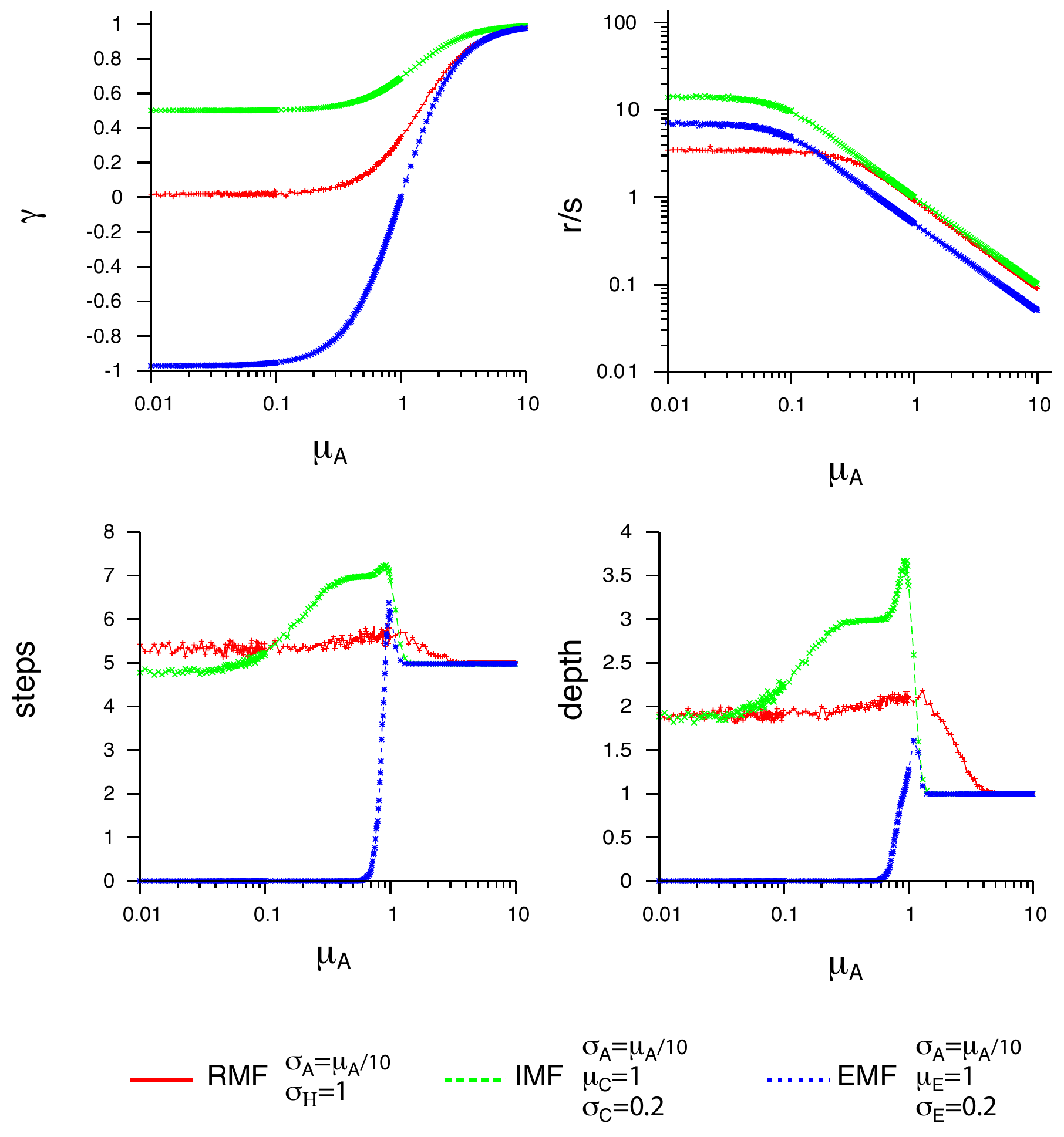}
\caption{Epistasis and chain  \co{measures} for different landscape models with L=5. The landscapes are build from an additive component with mean $\mu_a$ and variance $\sigma_a=\mu_a/10$ and an epistatic component. The epistatic component is: (RMF) an HoC model with $\sigma^2_{HoC}=1$; (IMF) an Ising model with mean incompatibility cost $\mu_c=1$ and variance $\sigma_c^2=0.1$; (EMF) an eggbox model with mean fitness effect $\mu_E=1$ and variance $\sigma_E^2=0.1$. We plot epistasis as measured by $\gamma$ and $r/s$, and chains as the number of chain steps and the maximum depth. }
\label{ChainsGamma} 
\end{center}
\end{figure}

\red{Furthermore, even though chain measures for other models (e.g. IMF and EMF) cannot be computed analytically, they can be retrieved from simulations (Figure \ref{ChainsGamma}).} Interestingly, several of the chain \co{measures} appear to be non-monotonic with respect to epistasis in the RMF model and even more so in the IMF model. Both the number of steps and the chain depth tend to have a maximum for intermediate values of epistasis, when the contributions of the additive component and of the interactions are comparable (Figures \ref{chain_stats} and \ref{ChainsGamma}). \co{Interestingly the type of interaction plays a important role in the structure of the chains, as only short chains are observed for the RMF and EMF models whereas long chains are observed for the IMF model. Furthermore, the size of the landscape also changes the dependence between epistasis \red{and chain measures (depth and abundance)} (Figure \ref{chain_stats}).}

%
%

\subsection{Generalized chains}

We can generalize the concept of chain by including cases where there is more than one fitness-increasing mutation from a given genotype, but all paths starting with these mutations eventually lead to the same genotype. 

We can define a \emph{generalized chain step} as a pair of genotypes $g\rightarrow g'$ such that all fitness-increasing paths from $g$ pass through $g'$. Note that in this case, a step can encompass one or more genotype(s) between $g$ and $g'$, whereas there is none in the original definition. A generalized chain can then be \red{defined as} a sequence of generalized chain steps. To compute the depth, \red{we} count all genotypes chains go through.

Generalized chains can also be interpreted in terms of constraints on evolution, but they are intrinsically non-local so they are not related to the out-degree distribution of the fitness graph. Instead, the total size (number of genotypes in the chain steps) of the generalized subchain ending at a genotype $g'$ corresponds to the size of the \emph{exclusive} basin of attraction of the genotype, \red{the set of genotypes that will inevitably evolve through $g'$}. Any path of increasing fitness starting at these genotypes will go through the genotype $g'$. 

This is particularly interesting for peaks, for which it captures their exclusive basins of attraction. \red{Non exclusive} basins of attraction are usually defined as genotypes from which the peak can be found \citep{Kauffman1993} and \red{their size is commonly} interpreted as a measure of the evolvability in the landscape. The exclusive basin of attraction of a peak measures how many genotypes have an obligatory end point, \textit{i.e.} how many genotypes are forced to evolve towards a single final state.

%
%

\section{Relations between  \co{measures}}

In this section we discuss the relations existing between the newly proposed  \co{measures} and the existing ones.

We expect that the new measures of epistasis $\gamma$ and $\gamma^*$ would be correlated to other measures of epistasis. In fact, we already discussed how they are related to some of the existing  \co{measures}. In particular, (i) for pairs of loci, $\gamma$ is directly related to the common definition of epistasis $e$ as $1-\gamma\propto e^2$: in fact, for the whole landscape we have $\gamma=1-\ev[e^2]/2\ev[s^2]$, while for a pair of mutations $\gamma_{i\rightarrow j}=1-\ev[e_{ij}^2]/2\ev[s_j^2]$; (ii) for the whole landscape, $\gamma$ can be rewritten as a function of the fitness correlation functions $\rho_d$ as $\gamma_d=(\rho_{d}-\rho_{d+1})/(1-\rho_{1})$; (iii) $\gamma^*$ is directly related to the number of square motifs with sign and reciprocal sign epistasis as $\gamma^*=1-\phi_s-2\phi_{rs}$.

Furthermore, it is possible to show that $\gamma$ is a function of the Fourier spectrum of the landscape, provided that the standard orthonormal basis is used for the Fourier series. If we denote by \co{$W_J$ the normalized weight of the coefficients of order $J$ in the Fourier spectrum, \textit{i.e.} the  sum of the squared coefficients of all $J$-loci interactions normalized by the sum of all squared coefficients, the relation is 
\beq
\gamma_d= 1-\frac{2\sum_{J=2}^{L}J\left[\sum_{\{m\ \mathrm{odd}\}}{d \choose m}{L-1-d \choose J-1-m} /{{L-1 \choose J-1}}\right]W_J}{\sum_{J=1}^{L}J W_J} \label{gammafourier}
\eeq
(see proof in the Appendix). Our measure of epistasis is therefore
\beq
\gamma=1-2\frac{\sum_{J=2}^{L}J(J-1)W_J}{\sum_{J=1}^{L}J(L-1)W_J}
\eeq 
which resembles another measure of \red{epistasis} $\sum_{J=2}^{L}W_J/\sum_{J=1}^{L}W_J$ \citep{Szendro2013}, showing again the close relation with previous measures of epistasis. The main difference is the weight of higher-order interactions: the contribution of $J$-loci interactions to $\gamma$ grows like $J^2$ for large $J$, so that the effect of interactions is stronger if they involve more loci.}

We also discussed how the number of chain steps, the number of peaks and sinks correspond to three different components of the out-degree distribution of the fitness graph. On the other hand, \red{we also suggested} that there is no direct relation between the amount of epistasis and the number of chains.

To evaluate in a more systematic way the relations between these and other  \co{measures}, we perform a correlation analysis similar to \citet{Szendro2013} but using models instead of experimental landscapes. We select the number of peaks, the number of sinks, the roughness/slope ratio (ratio between epistatic ``noise'' and additive component, see Appendix), $\gamma$ and $\gamma^*$ as  \co{measures} of epistasis, plus the number of chain steps and the maximum chain depth. We compute the Spearman correlation coefficients of all pairs of  \co{measures} in the RMF landscape model varying the model parameters (in particular, the ruggedness). 

\begin{table}
\begin{center}
\begin{tabular}{|c|ccccccc|}
\hline
              & peaks & sinks & $r/s$ & $\gamma$ & $\gamma^*$ & steps & depth \\ 
 \hline 
\multicolumn{8}{ |c| } {RMF model} \\
 \hline 
 peaks  & 1 & - & - &  -  & - & - & - \\
 sinks & 0.53 &1  & - & - & - & -   & - \\
 $r/s$ &  0.60  & 0.62 & 1 & - & - & -  & - \\
 $\gamma$ & 0.73 & 0.71 & 0.76 & 1   & - & -  & - \\ 
$\gamma^*$ & 0.77 & 0.73 & 0.71 & 0.88 & 1 & - & -  \\ 
 steps & 0.09 & 0.01  & 0.001 & $5 \cdot 10^{-4}$  & $4 \cdot 10^{-5}$ & 1   & - \\ 
 depth & 0.007 & 0.11 & 0.08 &  0.07 & 0.08 & 0.34 & 1  \\
 \hline 
\multicolumn{8}{ |c| } {IMF model} \\
 \hline 
 peaks  & 1 & - & - &  -  & - & - & - \\
 sinks & 0.69 &1  & - & - & - & -   & - \\
 $r/s$ &  0.72   & 0.45 & 1 & - & - & -  & - \\
 $\gamma$ & 0.73 & 0.46 & 0.99 & 1   & - & -  & - \\ 
$\gamma^*$ & 0.72 & 0.73 & 0.72 & 0.73 & 1 & - & -  \\ 
 steps & 0.25 & 0.01  &  0.09 & 0.09  & $2 \cdot 10^{-4}$ & 1   & - \\ 
 depth & $4 \cdot 10^{-5}$ & 0.08 & 0.02 &   0.02 & 0.20 & 0.68 & 1  \\  
 \hline 
\multicolumn{8}{ |c| }  {EMF model} \\
 \hline 
 peaks  & 1 & - & - &  -  & - & - & - \\
 sinks & 1.00 &1  & - & - & - & -   & - \\
 $r/s$ &  0.74   & 0.74 & 1 & - & - & -  & - \\
 $\gamma$ & 0.74 & 0.74 & 0.97 & 1   & - & -  & - \\ 
$\gamma^*$ & 0.99 & 0.99 & 0.74 &  0.74 & 1 & - & -  \\ 
 steps & 0.97 & 0.96  &  0.71 & 0.71  & 0.95 & 1   & - \\ 
 depth & 0.93 & 0.93 & 0.67 &   0.68 & 0.91 & 0.96 & 1  \\ \hline 
\end{tabular}
\caption{Spearman $\rho^2$ correlation of pairs of  \co{measures} across $10^4$ realizations of the RMF, IMF and EMF models with $L=5$, $\sigma_{HoC}=1$ (for RMF), $\mu_{I}=1$ and $\sigma_{I}=0.2$ (for IMF), $\mu_{E}=1$ and $\sigma_{E}=0.2$ (for EMF), $\sigma_a=\mu_a/10$ and $\mu_a$ log-uniformly distributed in [0.01,10]. These numbers correspond to the scale used in Figure \ref{ChainsGamma}. }
\label{table_stat}
\end{center}
\end{table}

The pairwise correlations (Table \ref{table_stat}) confirm the intuition that the  \co{measures} related to epistasis are all strongly correlated, the strongest correlation being between $\gamma$ and $\gamma^*$ \red{as expected}. \red{For RMF and IMF,} the correlation between these  \co{measures} and the chain measures, on the other hand, is especially low. \red{This is not the case for the EMF model.} This shows that the chain  \co{measures} quantify some landscape properties \red{that is  not simply correlated to the amount of epistasis.}

Interestingly, the  behavior of chains with epistasis is also apparent when we compare them across different models of epistatic interactions. In Figure \ref{ChainsGamma}, we show what happens in fitness landscapes with an additive contribution plus different epistatic models (HoC, Ising and Eggbox). The comparison between measures of epistasis and chain shows clearly that there is no simple relation between them, as all three models of interactions have different behavior. Chain  \co{measures} appear to depend strongly on the \co{nature} of the epistatic interactions, and therefore could provide useful, independent information. \red{The strong correlations between chain measures and epistasis observed for EMF relates to the observation that the chain measures are almost a two-steps constant function for EMF with a variable component of additive model.}

\section{\co{Measures} on two experimental landscapes}

\begin{figure}
\begin{center}
\includegraphics[scale=0.75]{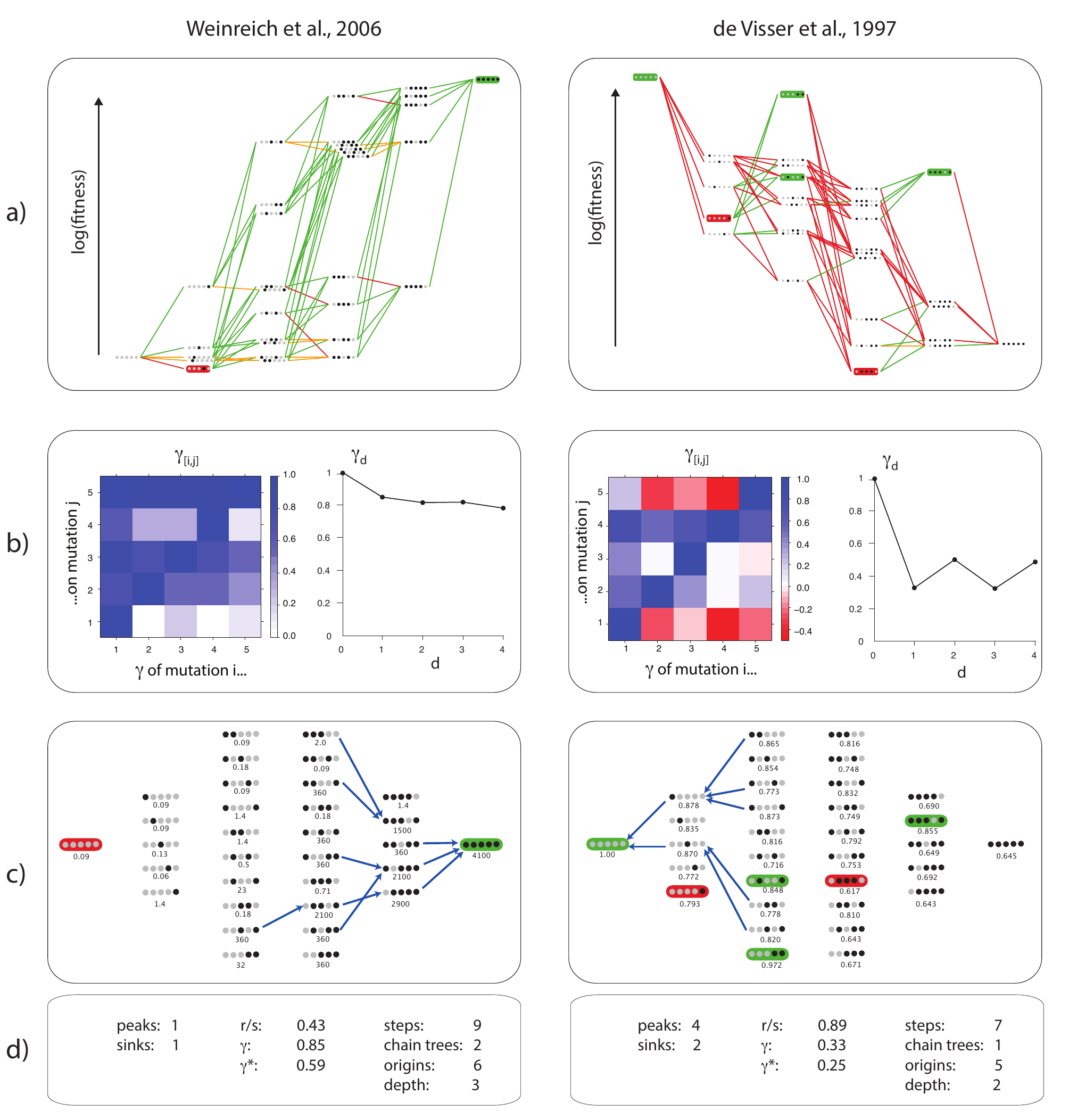}
\caption{Values of several  \co{measures} applied to two experimental landscapes. a) Illustrations of the landscapes using Magellan \citep{Magellan}. b) (left) Interactions between pairs of mutations $\gamma_{i\rightarrow j}$ : blue = no interaction, white = strong random interaction, red = strong interaction in sign; (right) Decay of $\gamma_d$ with Hamming distance. c) Chain steps in the landscape. d)  \co{Measures} for the landscape.}
\label{fig_gammaij} 
\end{center}
\end{figure}

As an example of application of the new  \co{measures} $\gamma$ and chains, we use them to analyze two complete experimental landscapes of size $L=5$. These landscapes are illustrated in Figure \ref{fig_gammaij}.

The first landscape is the  landscape of antibiotic (cefotaxim) resistance of $\beta$-lactamase mutations in an \textit{Escherichia coli} plasmid from \citet{Weinreich2006} (Figure \ref{fig_gammaij} left). The 5 mutations have a very strong effect that together give a $4\times10^4$ increase in antibiotic resistance and were therefore selected together. Given the huge selective advantage of the combined mutations, this landscape is single-peaked, where the peak corresponds to the \red{five-point} mutant. \red{It also has a single sink, that interestingly does not correspond to the wild type.}

The second is one of the \co{four} $L=5$ complete sublandscape (csI) \citep{Franke2011} of a larger landscape ($L=8$) of \co{deleterious mutations} 
in \textit{Aspergillus niger}  from \citet{deVisser1997} (Figure \ref{fig_gammaij} right). This landscape is a combination of unrelated deleterious mutations where epistatic interactions were not filtered by natural selection. This landscape has 4 peaks and 2 sinks; in fact, at present it is one of the most rugged among the completely resolved landscapes. 

As the \red{landscapes} were \red{derived} in completely different settings (co-selected beneficial for \red{$\beta$-lactamase} and random deleterious for \red{$Aspergillus$}), \red{we might not be surprised to find that these} landscapes \red{exhibit} very different structures. Indeed theoretical arguments support the intuition that  landscapes of co-selected mutations differ radically from landscape of random mutation \citep{Draghi2013,Greene2014,BlanquartArXiv}. The difference in ruggedness between \red{$\beta$-lactamase} and \red{$Aspergillus$} landscapes is  confirmed by the values of $\gamma$ (0.85 \textit{vs} 0.33) and $r/s$ (0.43 \textit{vs} 0.89).

To further explore the landscapes, we \co{compute the $\gamma_{i\rightarrow j}$ matrices} to illustrate and summarize the interactions between mutations (Figure \ref{fig_gammaij}b). In the \red{$\beta$-lactamase} landscape, there are some clear interactions between mutations (between the 2nd or the 4th and the 1st mutations, or between the 5th and the 4th) but none of these interactions is characterized by strong sign epistasis \red{(no red cell)}. On the other hand, the \red{$Aspergillus$} landscape contains several examples of interactions dominated by strong sign epistasis (for example, between the 2nd or  the 4th and the 1st or the 5th mutations).

Similar conclusions come from the analysis of the decay of $\gamma$ with distance. The decay in the \red{$\beta$-lactamase} landscape is immediate but decays slowly after the first mutation (Figure \ref{fig_gammaij}b left), resembling the behavior of RMF models. 

An interesting example of the power of $\gamma_d$ is represented by the \red{$Aspergillus$} landscape (Figure \ref{fig_gammaij}b right). This landscape shows a non-monotonic decay, with correlation $\gamma_d$ bouncing up and down. This indicates a clear compensatory structure of reciprocal sign epistasis, which is not only due to pairwise compensation, but extends to distance 4, \textit{i.e.} to the whole landscape. In fact, the behavior of $\gamma_d$ suggests a mixture of an RMF landscape and \co{an  extreme case of compensatory interactions, like the Eggbox model}. This surprising result does not come out in a straightforward way while looking at other  \co{measures}, \co{even when looking at} the Fourier spectrum \citep{Neidhart2013}. \red{Indeed, although the coefficient of the highest order in the Fourier decomposition measures the amount of eggbox, it compares to coefficients of smaller orders that have a complex intermingling when epistasis is not purely reciprocally signed at all orders. These coefficients contribute as well to the behaviour of $\gamma_d$.}

Finally, for both landscapes, the number of steps and the maximum depth of chains are higher than expected \red{theoretically}. For the \red{$\beta$-lactamase} landscape, the strength of epistasis suggests that this model should be close enough to an additive model that there should be a single chain of depth 1 and five steps ending at the peak, while we observe two chains with nine steps and their maximum depth is 3. This suggests that epistasis in this landscape is not random, but is structured in a way that constrains evolution and that is not captured by any of the models presented here. On the other hand, the number of steps and the depth of the single chain in the \red{$Aspergillus$} landscape are higher but not too different from the values expected for a RMF or EMF landscape, which is consistent with the above suggestion that this landscape resembles a mixture of RMF and EMF.

\section{Discussion}

In this work, we presented two new sets of  \co{landscape measures} which have a simple interpretation and cover a range of potential applications. \red{These measures among others have been implemented in MAGELLAN, a graphical tool to explore small fitness landscapes \citep{Magellan}}.

The first application is the measure of epistasis in a comparable way across landscapes. The correlation of fitness effects $\gamma$ is a natural  \co{measure} for this. This  \co{measure} can be used also for pairs of mutations, to explore the strength of epistatic interactions between mutations in a compact way. 

In terms of $\gamma$, there is a natural scale for the strength of epistatic interactions, from purely additive interactions ($\gamma=1$), \co{through} strong random interactions ($\gamma=0$) to a fully compensatory landscape ($\gamma=-1$). Interactions in landscapes with $\gamma<0$ are dominated by strong sign and reciprocal sign epistasis between most loci, therefore we expect such landscapes to be rare and possible only for some sets of mutations, \red{as selection tends to favor mutations with positive interactions}. In fact, the two experimental fitness landscapes analyzed have positive values of $\gamma$. Yet, the amount/strength of epistasis in the landscape by de Visser et al. is remarkably high: $\gamma=0.33$ means that \red{the fitness effect of a mutation in a given genotype, is a poor predictor of the fitness effect of the same mutation in a neighbor genotype that only differs by a \emph{single} mutation.}

Correlations of fitness effects are not only useful to quantify epistasis. Their decay $\gamma_d$ contains information on the nature of epistatic interactions and can reveal interesting signals. A clear example of that is the \red{$Aspergillus$} landscape studied here. The correlations $\gamma_d$ for this landscape show an oscillatory behaviour instead of the expected decay for random epistasis (\textit{i.e.} HoC like) or for incompatibilities (\textit{i.e.} Ising like), pointing towards a strong contribution of \co{``eggbox-like'' epistasis (reciprocal sign epistasis across multiple mutations)}. While the presence of pairwise reciprocal sign epistasis is not strange - it is actually quite common in compensatory interactions - the fact that reciprocal sign epistasis involves the whole landscape is quite surprising. In other words, starting from a first mutation chosen to be deleterious, it is not unreasonable that the second mutation could have a compensatory effect, but the mechanism behind the deleterious effect of the \emph{third} mutation and the compensatory effect of the \emph{fourth} mutation is obscure. It \co{relate to} complex  pathways of interactions at the molecular level.

\red{Many measures can only be computed if one has fitnesses for all combinations of the set of mutations (or subsets of)}. For example, the number of peaks lose meaning in a landscape with missing data, since the definition of a fitness maximum requires the knowledge of the fitness of all its neighbors. Since the fitness correlation functions $\rho_d$  can be computed even with missing data, the correlation of fitness effects can be estimated from equation (\ref{eq_rhod}) even for very sparse landscapes. The sparseness of the landscape could increase the error on the estimate, however this effect could be compensated by the larger size of the landscape. Landscapes containing a larger number of mutations would be also more representative of real gene or protein landscapes. 

\co{For some landscapes, only fitness ranks or the beneficial/deleterious nature of the fitness effects can be experimentally determined. Our measure $\gamma^*$ is appropriate for these landscapes. While $\gamma$ \red{depends not only on} positive and negative \red{epistasis}, but it is sensitive to its strength, $\gamma^*$ is based essentially on the fitness graph and therefore depends only on the sign of epistasis. $\gamma$ and $\gamma^*$ are strongly correlated across fitness landscape models. \red{Thus} a mismatch between $\gamma$ and $\gamma^*$ in real landscapes could point to some peculiar \co{nature} of epistatic interactions.}

\luca{Finally, the $\gamma$ and $\gamma_d$  \co{measures} could also be useful to estimate parameters of \red{theoretical} landscape models \red{from empirical data}, thanks to the availability of approximate analytical formulae for these quantities. For example, assuming that the underlying model of a landscape \red{is} the NK model, the  \co{measures} $\hat{K}=(L-1)(1-\gamma)$ is an approximately unbiased \co{Method-of-Moments} estimator of the parameter $K$, \textit{i.e.} $\ev[\hat{K}]\simeq K$ \red{(see eq. \ref{eqnk})}. A similar approach can be used for the parameters of other landscape models. The potential of these  \co{measures} for model inference and goodness-of-fit tests is yet to be studied. }

The other novel  \co{measures} that we proposed \red{are} the number of steps and depth of chains, \textit{i.e.} \red{mutations that are obligatory under the strong selection regime} in the landscape. These  \co{measures} have an immediate evolutionary interpretation, in terms of evolvability and constraints, yet they show peculiar properties.  \co{The most relevant one is that they are often non-monotonic in epistasis, as we have shown analytically. The lack of correlation between epistasis and chains shows that these new statistics can be used to obtain independent information about \emph{nature} of the interaction, instead of their strength.}

\red{Interestingly, the} number of chain steps, the chain depth \red{(Figure \ref{ChainsGamma})} and the total number of accessible paths \citep{Szendro2013} seem to be peaked at intermediate values of epistasis . All these \co{measures} are related both to evolvability  - deep chains show that fitness-increasing paths are open - and to constraints - chains  represent obligatory paths in evolution. This suggests an evolutionary interpretation of these peaks in terms of the tradeoff between evolvability (higher at low epistasis) and constraints (stronger with high epistasis). \co{Previous measures that were shown numerically and by some analytical approximations to be non-monotonic include the total number of accessible paths \citep{Szendro2013} and the number of exceedances, \emph{i.e}. the number of available fitness-increasing mutations after an evolutionary step \citep{NeidhartArXiv}, which are also related to evolvability and constraints. It is worth mentioning that chains and exceedance are both related to the out-degree distribution after one step of increasing fitness. This suggests that beyond the out-degree distribution, it is perhaps worth characterizing the sequences of out-degrees, e.g. the out-degree distribution along evolutionary paths.}

Real landscapes tend to have longer chains than expected according to \red{theoretical landscape models with random epistasis}. This is the case for both experimental landscapes studied here. This is especially apparent and interesting in the landscape by Weinreich et al., since it  \red{implies} some highly non-random structure of epistatic interactions for this set of mutations. This result has been found independently by randomization tests (Weinreich, personal communication) and cannot be seen in complex  \co{measures} of epistasis like the Fourier spectrum or the decay of correlations $\gamma_d$, however our chain  \co{measures} were able to capture this signal.

The notion of obligatory steps can be easily widen using the above definition of generalized chains. However, it is worth mentioning that the generalized chains say little about the local constraints in the fitness landscape. Quite on the contrary, they give information on exclusive basins of attraction for peaks. Therefore, typically, sizes of exclusive basins of attraction (generalized chains) are very informative about the \emph{convergence in state} \red{(similar endpoints regardless of the path)}. The larger the exclusive basins, the less uncertainty remains on the state, given a starting genotype. 
 Quite on the contrary, strict chains report local information of the landscape. They measure how path are constrained but do not necessarily predict what will be the final state. An interesting perspective would be to redefine the chains on the mutations themselves and not on the genotypes (e.g. mutation at locus $i$ is always followed by a mutation at locus $j$, independently of the genotype). Quite clearly, strict chains are informative about the \emph{convergence in path} in the landscape \red{(similar path regardless of the endpoints)} rather than the convergence in state.

Chains are natural  \co{measures} both from the evolutionary point of view and from the mathematical point of view, since they are related to the out-degree distribution of the fitness graph. For the small complete landscapes currently available, the number of peaks, chains and sinks summarize most of the information present in the out-degree distribution. However, for larger landscapes or for incomplete ones, other components of the out-degree distribution (or other properties, like its variance) could be useful as \co{measures} for a finer characterization of fitness landscapes. 

We are still far from predicting evolution on real landscapes based on their \co{measures}, partly because of the incomplete knowledge of the structure of real landscapes, and partly because of the lack of  \co{measures} with a natural evolutionary interpretation. In the future, we expect to witness a strong increase in the number of \red{published empirical} landscapes that will be experimentally resolved.  The  \co{measures} that we propose here will therefore find applications in the understanding and classification of these landscapes, as well as in studies of model landscapes. The correlation of fitness effects is a natural measure of epistasis that is comparable across landscapes, while the decay of correlations with mutation distance and the new chain  \co{measures} will be useful tools to discriminate and classify these landscapes. Chains also highlight the interplay of constraints and evolvability that influence evolution on complex landscapes.   

\section{Acknowledgments}

We thank Joachim Krug, Ivan Szendro, Johannes Neidhart, Arjan de Visser, Nick Barton, RA Watson for useful comments and discussions. The work was funded by grant ANR-12-JSV7-0007 from Agence Nationale de la Recherche. BS acknowledges support from the Bonn-Cologne Graduate School.

%
%
\bibliographystyle{evolution}
\bibliography{ferretti_etal}


\appendix

\section{Appendices}

\co{\subsection{Common landscape  \co{measures}} \label{sec_stat}
\red{In this section we present some common landscape measures that can be applied as statistics for experimental landscape data. Notation:} a genotype $g$ is a sequence of alleles $g=(A_1A_2A_3\ldots A_L)$ of length $L$. For biallelic landscapes, $A_i\in\{0,1\}$ and $S_i=2A_i-1$. 

Some of the most common  \co{measures} for fitness landscapes $f(g)$ are:
\begin{itemize}
\item number of peaks \citep{Weinberger1991}: it is the number of genotypes such that all their neighbours have lower fitness, \textit{i.e.} the number of local fitness maxima.
\item $r/s$ (roughness/slope) ratio \citep{Aita2001}: the landscape is fitted to a linear model (a linear combination of $A_i$s plus a constant) by least squares. The slope $s$ is the average modulus of the coefficients of the $A_i$s. The roughness $r$ is the quadratic mean of the residuals of the regression. The measure of epistasis is their ratio $r/s$.
\item \red{fraction of epistatic interactions} \citep{Weinreich2005a,Poelwijk2007}: the fraction of all pairs of mutations from all possible genotypes that show magnitude, sign or reciprocal sign epistasis.
\item number of accessible paths \citep{Weinreich2006}: assume that the absolute fitness maximum corresponds to the genotype $g=(111\ldots 1)$. Count the number of paths of mutations $0\rightarrow 1$ starting from $g=(000\ldots 0)$ to $(111\ldots 1)$ such that fitness increases after each mutation. This is the number of direct accessible paths to the maximum from its antipodal genotype.
\item Fourier expansion and spectrum \citep{Stadler1996,Weinreich2013,Neidhart2013}: 
The coefficients $a_{i_1\ldots i_J}$ of the Fourier expansion are uniquely defined in terms of the Fourier decomposition 
\beq
f(g)=f_0+\frac{1}{2^{N/2}}\sum_{J=1}^{L}\sum_{\{i_1\ldots i_J\}}a_{i_1\ldots i_J}S_{i_1}\ldots S_{i_J}
\eeq
where $\{i_1\ldots i_J\}$ are ordered sets. The Fourier spectrum is defined by the sum of squared coefficients for interactions of $J$ \co{loci}: $B_J=\sum_{\{i_1\ldots i_J\}}a^2_{i_1\ldots i_J}$. Epistasis is usually measured by $\sum_{J\geq 2} B_J/\sum_{J\geq 1} B_J$.
\end{itemize}
More details can be found in the review by \citet{Szendro2013}.

}

\co{\subsection{Models of fitness landscapes} \label{sec_models}
In this section we briefly illustrate some common models of fitness landscapes that will be used in this study. Most of them are illustrated in Figure \ref{models}. Please note that we only considered here models of $L$ biallelic loci. 
A mathematical  formulation of these models is given in the next section.

\begin{figure}[ht]
\begin{center}
\includegraphics[scale=0.85]{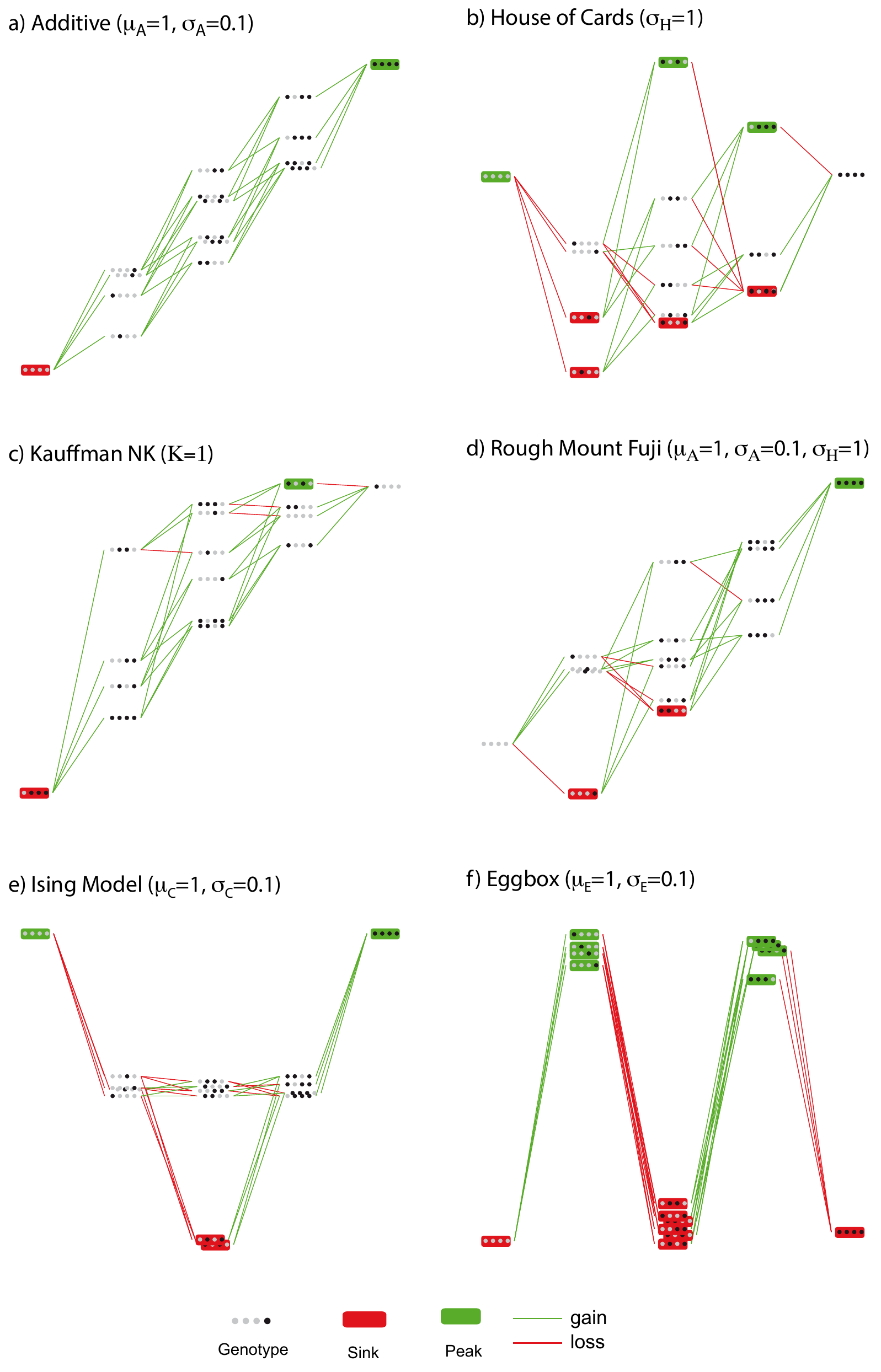}
\caption{ {\bf Models of fitness landscapes}. Realizations of random landscapes obtained from the models discussed in the introduction, using Magellan \citep{Magellan}.}
\label{models} 
\end{center}
\end{figure}

\subsubsection{The Additive model (a.k.a. multiplicative model)}

This is a model for non-interacting mutations with independent fitness effects. The fitness is simply the product of the fitness contributions of each locus: fitness effects of different mutations are multiplied. In log-scale, this corresponds to summing the fitness effect of each mutation; for this reason this models is called ``additive''. Here, the fitness effects of different mutations are randomly drawn from a Gaussian distribution with mean $\mu_a$ and variance $\sigma_a^2$. As there an independent contribution of each locus, the dimension of interaction is $1$ (since each locus ``interacts'' only with itself).

In terms of Fourier decomposition, in this landscape all coefficients of second order and higher are zero.

\subsubsection{The House-of-Cards (HoC) model}

This is a model for random, uncorrelated fitness landscapes \citep{Kingman1978}. The fitness of each genotype is independent on the fitnesses of other genotypes. Here, it is randomly drawn according to a Gaussian of mean 0 and variance $\sigma_{HoC}^2$. As this models corresponds to full interaction between the loci, the dimension of interaction is $L$.

In terms of Fourier decomposition, the coefficients are random variables with a marginal Gaussian distribution centered in 0.

\subsubsection{The Rough Mount Fuji (RMF) model}

This model interpolates between additive and uncorrelated fitness landscapes by adding the two \citep{Aita2000}. The fitness is computed as the sum of an additive contribution and a HoC contribution. Here, the model is tuned by three parameters: mean $\mu_a$  and variance $\sigma_a^2$ for the additive part and variance $\sigma_{HoC}^2$ for the HoC part. (In the literature, this model is often defined with constant additive fitness effects, \textit{i.e.} $\sigma_a^2=0$). The model converges to an additive model when $\sigma_{HoC}^2\ll \mu_a^2+\sigma_a^2$ and to a HoC model when $\sigma_{HoC}^2\gg \mu_a^2,\sigma_a^2$. The dimension of interactions is a mixture of dimension $1$ and dimension $L$.

The Fourier decomposition is a linear function of the landscape, so it is a combination of the additive and the HoC decompositions.


\subsubsection{The NK model}

This landscape model with $N=L$ loci interpolates between additive and uncorrelated fitness landscapes by combining uncorrelated fitness contributions (\textit{i.e.} small HoCs) from $L$ groups of  $K+1$ loci  in an additive way \citep{Kauffman1989}.
\red{There are different ways to choose the groups of interacting loci and while several properties such as mean
number and mean height of local optima depend only weakly on the particular choice made \citep{Weinberger1991}, others seem to behave quantitatively different for some interaction choices \citep{Schmiegelt2014}. Nonetheless it
has been shown that the fitness correlation function is strictly independent of the interaction
choice \citep{Campos2002} and consequently $\gamma$ does not depend on it either, while the
number of chains may still be influenced by it. The number of interacting loci is $K+1$ and the interpolation is controlled by the parameter $K\in\{0,1\ldots L-1\}$: $K=0$ corresponds to an additive model with independent contributions from each locus, while $K=L-1$ corresponds to an HoC model.} The dimension of interaction is $K+1$.

\subsubsection{The Ising and the IMF models}

This model originates from statistical physics \citep{Mezard1987}, but has an immediate interpretation in terms of pairwise allele incompatibilities. In this model, each pair of interacting loci with different alleles causes a reduction in fitness. Here, loci interact only if they are neighbors in the genotype sequence (locus $i$ interacts only with locus $i-1$ and $i+1$) and the first and the last locus have a single interaction (loci are arranged on a string). The fitness cost for each pair is drawn from a Gaussian with mean $\mu_c$ and variance $\sigma_c^2$. More general models based on allelic incompatibilities correspond to the Sherrington-Kirkpatrick model  and other spin glass models in statistical physics \citep{Mezard1987}. The dimension of interaction is $2$ as interactions only occurs between pairs. We also combined incompatibility interactions (Ising model) with an independent fitness contribution (additive model) in an ``Ising Mount Fuji'' (IMF) model in the same way the RMF is set.

In terms of Fourier decomposition, in this landscape all coefficients of third order and higher are zero.

\subsubsection{The Eggbox  and the EMF models}

This model represents the extreme example of reciprocal sign epistasis of highest dimension. In this model, all genotypes in the landscapes have either low or high fitness. All the neighbours of a high-fitness genotype have low fitness, and vice versa. Therefore, in this landscape, each mutation is either deleterious (from high to low fitness) or  compensatory (from low to high fitness). Fitnesses are given by a Gaussian with mean $f_0\pm\mu_E/2$ and a small variance $\sigma_E^2$. The dimension of interactions in this landscape is $L$. We also combine the eggbox interactions with independant contributions in an ``Eggbox Mount Fuji'' (EMF) landscape that is built like the RMF or the IMF.

In terms of Fourier decomposition, this landscape is dominated by the contribution of the $L$th-order term (that is, the term of highest order).
}

\subsection{Formulae for $\gamma_d$ in model landscapes}

We assume an observed fitness $f(g)$ given by the model 
\beq
f(g)=\sum_{i=1}^L\mu_iA_i+f_e(g)+\varepsilon_g
\eeq
that is, an additive contribution $\sum_{i=1}^L\mu_iA_i$, an epistatic contribution $f_e(g)$ from some fitness model, plus the effect of measurement errors $\varepsilon_g$. We assume these errors to be unbiased 
and uncorrelated:  $\ev[\varepsilon_g]=0$, $\cov[\varepsilon_g,\varepsilon_{g'}]=\delta_{gg'}\sigma_g^2$ where $\delta_{gg'}$ is the Kronecker delta, \textit{i.e.} $\delta_{gg'}=1$ when $g=g'$ and 0 otherwise.

We define the mean squared additive effect $\mu^2=\sum_{i=1}^L\mu_i^2/L$ and the mean squared experimental error $\sigma_\varepsilon^2=\sum_{g}\sigma_g^2/2^L$.

The expected value of $\gamma_d$ can be computed approximately by taking the ratio of the expected values of numerator and denominator of eq. (\ref{eq_gammad}) rearranged as eq. (\ref{def_gamma2})\red{, instead of the expected value of the ratio (the $\simeq$ sign in all our formulae refers to this approximation)}. The result is
\beq
\ev[\gamma_d]\simeq 1-\frac{\ev[(s_e(g)-s_e(g_d))^2]+4\sigma_\varepsilon^2}{2\mu^2+2\ev[(s_e(g))^2]+4\sigma_\varepsilon^2}
\eeq
with $s_e(g) = f_e(g_{[j]}) - f_e(g)$ being the analogue of $s(g)$ restricted to the epistatic contribution. Note that $\ev[(s_e(g)-s_e(g_d))^2] = 2(1-\ev[ {\gamma_e}_d]) \ev[ (s_e(g))^2]$ if ${\gamma_e}_d$ is the $\gamma_d$ statistic of the epistatic contribution $f_e(g)$.

%

\subsubsection{Additive model}
In these models $f_e=0$ and the only reduction in correlation is due to experimental noise:
\beq
\ev[\gamma_d]\simeq 1-\frac{2\sigma_\varepsilon^2}{\mu^2+2\sigma_\varepsilon^2}
\eeq

\subsubsection{RMF and HoC models}
In these models, $f_e(g)$ corresponds to the HoC model, \textit{i.e.} they are i.i.d. random variables. 
 Denote by $\sigma_{HoC}^2$ the variance of the distribution of fitnesses in the HoC model:
\beq
\ev[\gamma_d]\simeq 1-\frac{2\sigma_{HoC}^2+2\sigma_\varepsilon^2}{\mu^2+2\sigma_{HoC}^2+2\sigma_\varepsilon^2}
\eeq
Since $\mu^2=\mu_a^2+\sigma_a^2$ for a Gaussian distribution of additive fitness effects, we obtain equation (\ref{eqhoc}) for $d=1$ and $\sigma_\varepsilon^2=0$.

\subsubsection{NK models}
In these models, $f_e(g)=\sum_{i=1}^L F_i$ where the $F_i$s are i.i.d. random variables that depend on $i$ and other $K$ indices, randomly chosen. The $F_i$s have mean $f_0$ and variance $\sigma_{NK}^2$.

The fitness correlation function $\rho_d$ is known exactly for the pure NK model \citep{Campos2002}: $\rho_d = \frac{(L-K-1)!(L-d)!}{L!(L-K-d-1)!}$. With eq. (\ref{eq_rhod}) it is straightforward to compute $\ev[ {\gamma_e}_d]$ from this.
The variance $\ev[ s_e(g)^2]$ is given by $2(K+1)\sigma_{NK}^2$ because on average $K+1$ of the $F_i$ change in a single mutation, each of these differences having twice the variance of the fitness contribution (since the variance of the difference of two i.i.d. variables is twice their variance). The final result is
\beq
\ev[\gamma_d]\simeq 1-\frac{2(K+1)\sigma_{NK}^2\left(1-\frac{{L-1-d \choose K} }{{L-1 \choose K}}\right)+2\sigma_\varepsilon^2}{\mu^2+2(K+1)\sigma_{NK}^2+2\sigma_\varepsilon^2}
\eeq
Substituting $\mu^2=\sigma_\varepsilon^2=0$ and $d=1$ yields equation (\ref{eqnk}).

\subsubsection{Ising model}


In the following, we define $S_i=2A_i-1\in\{-1,+1\}$.
Using the above notations, in these models, $f_e(g)=\sum_{i}J_{i}S_iS_{i+1}$ where the incompatibility coefficients $J_i$ are randomly extracted from a Gaussian distribution with mean $\mu_c$ and variance $\sigma_c^2$. We define $J^2=\ev[J_{i}^2]=\mu_c^2+\sigma_c^2$.

A mutation at locus $i$ will invert contributions of the terms containing $J_i$ and $J_{i-1}$ adding $8J^2$ to $\ev[ (s_e(g))^2]$. At the edge of the genome, loci interacts with only one neighbor, so this is reduced by a factor of 2, thus $\ev[(s_e(g))^2] = \frac{8(L-1)J^2}{L}$.
  
Only mutations at $j-1$ and $j+1$ affect the value of $s_j(g) - s_j(g_d)$. Choosing $d$ mutations out of $L-1$ and applying the hypergeometrical distribution, there are probabilities $\frac{\binom{L-3}{d-2}}{\binom{L-1}{d}}$ and
$\frac{2\binom{L-3}{d-1}}{\binom{L-1}{d}}$ to choose both or exactly one of them, respectively. Each relevant mutation changes the effect of mutating $j$ by $\pm4J$. Thus (ignoring boundaries) 
 $\ev[(s_e(g)-s_e(g_d))^2]
 \approx 16J^2\frac{2\binom{L-3}{d-2}+2\binom{L-3}{d-1}}{\binom{L-1}{d}} = 32J^2\frac{d}{L-1}.
$ 
On the boundary only one mutation can influence $s_j$, resulting in a reduced contribution of $16J^2\frac{d}{L-1}$ and so together:
\begin{align*}
 \ev[ (s_e(g)-s_e(g_d))^2 ] &= 32J^2\frac{d}{L-1}\frac{L-1}{L} = 32J^2\frac{d}{L}
\end{align*}
and therefore the result is
\beq
\ev[\gamma_d]\simeq 1-\frac{16dJ^2/L+2\sigma_\varepsilon^2}{\mu^2+8(L-1)J^2/L+2\sigma_\varepsilon^2}
\eeq
%
%

\subsubsection{Eggbox model}
In this model, $f_e(g)=f_0+\frac{\mu_E}{2}\prod_{i=1}^L (-1)^{A_i}$, i.e each mutation switches the fitness from the highest value ($f_0+\mu_E/2$) to the lowest ($f_0-\mu_E/2$) or the other way. The difference in the epistatic fitness effects from two genotypes separated by an odd number of mutations  is $\pm\mu_E$, while it is 0 from two genotypes separated by an even number of mutations, therefore
\beq
\ev[\gamma_d]\simeq 1-\frac{\mu_E^2\left(1-(-1)^d\right)+2\sigma_\varepsilon^2}{\mu^2+\mu_E^2+2\sigma_\varepsilon^2}
\eeq

\subsection{Formulae for chains in RMF}

We consider a RMF model with equal additive fitness contribution $s$
for each mutation, in addition to an HoC model with distribution
$p(f)$ and cumulative distribution $C(f)=\int_{-\infty}^fp(x)dx$.

Sorting the genotypes in order of their additive fitness, there are
${L\choose k}$ genotypes at the $k$th level with $L-k$ mutations with
positive additive fitness effect and $k$ with negative one. The
average number of chain steps can be obtained as the sum over all
possible steps of the probability of being a chain step. It is
\begin{align}
\#\mathrm{steps}&=\sum_{k=0}^L{L \choose k} \left[k \int dx\,p(x)
(1-C(x+s))C(x+s)^{k-1}C(x-s)^{L-k}\right.+ \nonumber \\
+ & \left. (L-k) \int dx\,p(x) (1-C(x-s))C(x-s)^{L-k-1}C(x+s)^{k}
\right] = \nonumber \\
=&L \int dx\,p(x) [2-C(x-s)-C(x+s)]\cdot [C(x+s)+C(x-s)]^{L-1}
\end{align}

The number of origins and of \co{chain trees} can be obtained in a similar way
using the Cayley tree/Bethe lattice approximation. In this framework,
we approximate locally the hypercube by a Cayley tree, \textit{i.e.} a tree
with L branches at each node. This means that we neglect the
overlapping between the next-to-nearest neighbours of the genotype
considered and we assume them to be $(L-1)^2$ independent genotypes
instead of $L(L-1)/2$.

The probability that a genotype is the origin of a chain is product of
the probability of having out-degree 1 and that all the other
neighbours of lower fitness have out-degree different than 1:
\begin{align}
\#\mathrm{origins}&=\sum_{k=0}^L{L \choose k} \biggl\{(L-k) \int
dx\,p(x) (1-C(x-s))\cdot(C(x-s)-p_+(x,s,k))^{L-k-1} (C(x+s)-p_-(x,s,k))^{k}+ \nonumber \\
+& k\int dx\,p(x) (1-C(x+s))\cdot(C(x+s)-p_-(x,s,k))^{L-k} (C(x-s)-p_+(x,s,k))^{k-1} \biggl\}
\end{align}
where we define the probabilities that a neighbour genotype is the starting or ending point of a chain step ending at level $k$:
\begin{align}
p_+(x,s,k)&=\int^{x-s}dy\,p(y)C(y+s)^kC(y-s)^{L-k-1}\\
p_-(x,s,k)&=\int^{x+s} dy\,p(y)C(y+s)^{k-1}C(y-s)^{L-k}
\end{align}

The probability that a genotype is the endpoint of a chain is the
difference between the probability of being the final genotype of a chain step and
the probability of being an intermediate point in a chain. The results
are
\begin{align}
\#\mathrm{endpoints}&=\sum_{k=0}^L{L \choose k} \biggl[1-\int dx\,p(x)
\left(
1-p_+(x,s,k)\right)^{L-k} 
\cdot \left(
1-p_-(x,s,k)\right)^k\biggr] \\
\#\mathrm{intermediates_\uparrow}(s)&=\sum_{k=0}^L{L \choose k} (L-k)
\int dx\,p(x) (1-C(x-s)) \cdot \biggl[ C(x-s)^{L-k-1}C(x+s)^{k}+ \nonumber
\\
-& \left(
C(x-s)-p_+(x,s,k)\right)^{L-k-1} 
\cdot \left(
C(x+s)-p_-(x,s,k) \right)^{k} \biggl]
\end{align}
\begin{align}
\#\mathrm{intermediates}&=\#\mathrm{intermediates_\uparrow}(s)-\#\mathrm{intermediates_\uparrow}(-s)
\\
\#\mathrm{\co{chain\ trees}}&=\#\mathrm{endpoints}-\#\mathrm{intermediates}\end{align}
These formulae could be further simplified if $p(f)$ is the Gumbel
distribution \citep{NeidhartArXiv}.

\subsection{Proofs}
  \subsubsection{Relation between $\gamma$ and type of epistasis}
 The most extreme values of $\gamma$ for different types of epistasis are obtained in the case $L=2$. 
Denote by $f_{00},f_{10},f_{01},f_{11}$ the log-fitness values. The function $\gamma$ is defined as
\begin{equation*}
\gamma=\frac{2[(f_{11}-f_{10})(f_{01}-f_{00})+(f_{11}-f_{01})(f_{10}-f_{00})]}{(f_{11}-f_{10})^2+(f_{01}-f_{00})^2+(f_{11}-f_{01})^2+(f_{10}-f_{00})^2}
\end{equation*}
Since $\gamma$ is a correlation, $-1\leq \gamma\leq 1$. $\gamma$ is a continuous function and it is also invariant under permutations of loci and alleles, so from now on we will restrict to the subspace with $f_{00}\leq f_{01}\leq f_{10}$. Each of the three partitions of this subspace corresponding to magnitude, sign and reciprocal sign epistasis is connected, therefore the image of each one of them under $\gamma$ is an interval. 

By definition, magnitude epistasis results in $\gamma\geq 0$ since all fitness jumps have the same sign. The extreme values 0 and 1 are both realized: $\gamma=0$ in landscapes with $f_{00}=f_{01}=f_{10}<f_{11}$, while $\gamma=1$ in landscapes with $f_{11}=f_{10}+f_{01}-f_{00}$. Therefore, for magnitude epistasis, $0\leq \gamma \leq 1$.

Reciprocal sign epistasis require that all fitness jumps change in sign after a mutation, therefore results in $\gamma<0$ by definition. The extreme case $\gamma=-1$ is realized in the landscape $f_{01}=f_{10}>f_{00}=f_{11}$, while the other extreme case $\gamma\rightarrow 0$ is realized for the landscapes with
$f_{00}<f_{01}=f_{10}>f_{11}$ for $(f_{00} - f_{01}) \to 0$. Therefore, for reciprocal sign epistasis, $-1\leq \gamma < 0$.

Finally, sign epistasis can have both signs of $\gamma$. It is easy - although tedious - to show that there are no critical points of $\gamma$ inside the space of landscapes with sign epistasis (with $L=2$), therefore the extremal values should appear on the border. There are essentially two borders: $(f_{11} - f_{10}) \to 0$ and $f_{11}=f_{01}$. On the first border, we have $\gamma>0$ and the upper limit is $\gamma\rightarrow 1$ for the landscapes with $f_{00}=f_{01}<f_{10}$. On the second border, we have $\gamma=-\frac{2(f_{10}-f_{11})(f_{11}-f_{00})}{(f_{11}-f_{10})^2+(f_{10}-f_{00})^2+(f_{11}-f_{00})^2}$ that reaches the minimum value $\gamma=-1/3$ at the landscapes with $f_{10}-f_{11}=f_{11}-f_{00}$ (imposing the derivative with respect to $f_{10}$ to be null).  Therefore, for  sign epistasis, $-1/3\leq \gamma < 1$.

  \subsubsection{Relation (\ref{eq_rhod}) between $\gamma_d$ and the fitness correlation function}
  Denote the fitness correlation function at distance $d$ by $\rho_d=\cor[ f(g),f(g_d)]$. We use the identity 
  \begin{eqnarray*}
  s_j(g)s_j(g_{[i_1i_2\ldots i_d]}) & =  & f(g_{[j]})f(g_{[ji_1i_2\ldots i_d]})-f(g)f(g_{[ji_1i_2\ldots i_d]}) \\
  & &-f(g_{[j]})f(g_{[i_1i_2\ldots i_d]})+f(g)f(g_{[i_1i_2\ldots i_d]}) 
  \end{eqnarray*}
  Averaging over the all mutations and all genotypes, then dividing by $\var[f(g)]$, the above terms give rise to:
\begin{eqnarray*}  
    \ev[s _j(g)s_j(g_{[i_1i_2\ldots i_d]}) ] & = & 2 ( \ev[f(g)f(g_d)] - \ev[f(g)f(g_{d+1})]  ) \\
                                                                 & = & 2 ( \cov[f(g),f(g_d)] - \cov[f(g),f(g_{d+1})]  ) \\
                                                                 & = &  2(\rho_d-\rho_{d+1})\var[f(g)]
\end{eqnarray*}  
Summing over genotypes and mutations, the numerator of (\ref{eq_gammad})
  becomes $2^L L{L-1 \choose d} 2(\rho_d-\rho_{d+1})\var[f(g)]$. The denominator of (\ref{eq_gammad}) can be computed in a similar way by choosing $d=0$, obtaining ${L-1 \choose d}2^L L \cdot 2(\rho_0-\rho_{1})\var[f(g)]$. $\gamma_d$ is their ratio 
\begin{equation*}
\gamma_d=\frac{\rho_d-\rho_{d+1}}{\rho_0-\rho_{1}}
\end{equation*}
and since $\rho_0=\cor[ f(g),f(g)]=1$, we obtain the result (\ref{eq_rhod}).

\subsubsection{Relation (\ref{gammamotifs}) between $\gamma^*$ and the fractions of square motifs}
We consider a landscape without ties (\textit{i.e.} all genotypes have different fitness values). In this case, $(s^*_j)^2=1$ and therefore $\gamma^*=\frac{1}{2^LL(L-1)}\sum_g\sum_{i,j\neq i} (s_j^*(g)\cdot s_j^*(g_{[i]}))=\ev[s^*(g)\cdot s^*(g_1)]$ where the average is over all genotypes and pairs of mutations, or equivalently over all square motifs and over their sides. The average over the two sides of a motif is $\ev[s^*(g)\cdot s^*(g_1)]=(1+1)/2=1$ with magnitude epistasis, $(1-1)/2=0$ with sign epistasis and $(-1-1)/2=-1$ with reciprocal sign epistasis. To obtain the global average, we multiply these results by the fraction of motifs of each kind, \textit{i.e.} $\gamma^*=1\cdot \phi_m+0\cdot \phi_s-1\cdot \phi_{rs}$. Since they sum to 1, we have $\phi_m=1-\phi_s-\phi_{rs}$ and substituting it we obtain the relation (\ref{gammamotifs}).

\subsubsection{Number of chain steps in the HoC model}
The average number of chain steps is given by the number of mutations in the landscape, $L\cdot 2^L$, multiplied by the probability that the mutation is a chain step. This is the probability that among the initial genotype and all its neighbours, the final one is the most fit and and the initial one is second in the fitness rank. Since fitness values are random and uncorrelated, the probability that a value is maximum among $L+1$ is $1/(L+1)$ and the conditional probability that another value is second is $1/L$, therefore we have that the average number of steps is $L\cdot 2^L/L(L+1)=2^L/(L+1)$.

\subsubsection{Relation (\ref{gammafourier}) between $\gamma$ and the Fourier spectrum}
We define $\bar{n}(J,d)=\lfloor\min((J-2)/2,(d-1)/2)\rfloor$. Since the Fourier basis is orthonormal, each component of the spectrum gives an independent contribution to the numerator and denominator of $\gamma$. The contribution of each $B_J$ to the denominator of (\ref{def_gamma2}) is $4J$ since there are $J$ nonzero mutations with fitness effect $\pm 2a_{i_1\ldots i_J}$ each. For the numerator,  there is again a factor $J$ contributions (from nonzero mutations) multiplied by the square of the fitness effect of each term (averaged over the choice of the other $d$ mutations). The fitness effect is $\pm 4a_{i_1\ldots i_J}$ if an odd number of the $d$ mutations lie within the indices ${i_1\ldots i_J}$, and 0 otherwise. The probability that this number is odd is given by the sum of odd terms of the hypergeometric distribution with parameters $d,J-1,L-1$, therefore we have
\beq
\ev[\gamma_d]\simeq 1-\frac{2\sum_{J=2}^{L}JB_J\left(\sum_{n=0}^{\bar{n}(J,d)}
\frac{{d \choose 2n+1}{L-1-d \choose J-2n-2}}{{L-1 \choose J-1}} \right)}{\sum_{J=2}^{L}J B_J}
\eeq
and since $W_J=B_J/\sum_{I=1}^N B_I$, we have the equation (\ref{gammafourier}).

\end{document}